# Active labor market policies for the long-term unemployed: New evidence from causal machine learning

Daniel Goller[1,2], Michael Lechner[1], Tamara Pongratz[3], Joachim Wolff[3]*

[1]Swiss Institute of Empirical Economic Research, University of St. Gallen
[2]Centre for Research in Economics of Education, University of Bern
[3]Institute for Employment Research, Nuremberg

This version: May 2023 (first version: August 2021)

**Abstract:** Active labor market programs are important instruments used by European employment agencies to help the unemployed find work. Investigating large administrative data on German long-term unemployed persons, we analyze the effectiveness of three job search assistance and training programs using Causal Machine Learning. Participants benefit from quickly realizing and long-lasting positive effects across all programs, with placement services being the most effective. For women, we find differential effects in various characteristics. Especially, women benefit from better local labor market conditions. We propose more effective data-driven rules for allocating the unemployed to the respective labor market programs that could be employed by decision-makers.

**Address for correspondence**: Daniel Goller, Centre for Research in Economics of Education, University of Bern, Schanzeneckstrasse 1, CH-3001 Bern, Switzerland, Daniel.Goller@unibe.ch

* Michael Lechner is affiliated with CEPR, London, CESifo, Munich, IAB, Nuremberg, IZA, Bonn, and RWI, Essen. Joachim Wolff is affiliated with the Ludwig-Maximilian University, Munich, LASER, Nuremberg, and GLO, Essen. Disclosure statement: Support of the Swiss Science Foundation (grant SNST 407540_166999) and of the IAB under a grant for the project "Estimating heterogeneous effects of the schemes for activation and integration on welfare recipients' outcomes: Enhanced analyses by the application of machine learning algorithms" is gratefully acknowledged. A previous version of the paper was presented at the CML Workshop, 2020, St. Gallen; EALE, 2021, Padova (online); VfS annual meeting 2021, Regensburg (online), ESPE, 2022, Calabria. We thank the participants for their helpful comments and suggestions. The usual disclaimer applies.

# 1 Introduction

Bringing means-tested benefit recipients back to work is among the most challenging tasks for employment agencies. Still, it is of high interest for all, society, the state, and most importantly, the unemployed, to increase their chances of finding appropriate jobs and leave long-term unemployment. The classical approach of employment agencies in many industrialized countries is to provide active labor market programs (ALMP), such as job search and training programs, to selected unemployed individuals. Therefore, it is of public interest to understand whether those ALMPs are beneficial for the long-term unemployed and which program works best, and for which types of unemployed persons. Moreover, learning how to allocate programs effectively is key from a policy perspective.

This study shares the interest of policymakers in gaining a better understanding of the employment agencies' tools by evaluating the existing ALMP and finding ways to improve the counseling process and allocation of ALMP to the long-term unemployed. The literature on active labor market policy evaluation is substantial (e.g., the meta-studies of Kluve, 2010; Card et al., 2010, 2018; or less recent surveys on ALMP evaluation in Heckman et al., 1999; Martin and Grubb, 2001) and agrees on negative lock-in effects on employment outcomes in the short-run, while in the medium and long-run effects are often positive but vary substantially. Differential impacts of ALMPs on men and women are frequently studied and documented, for example, by Card et al. (2018), who report larger positive impacts for women in ALMP. More systematically, Bergemann and van den Berg (2008) review previous ALMP studies focusing on gender-specific effects of ALMP and report larger positive employment effects for women than for men only when female labor market participation is low.[1] For studies investigating the long-term unemployed, the meta-study of Card et al. (2018) finds larger positive employment effects of ALMP reported than in studies investigating short-term unemployed.

Previous empirical evaluations of ALMP in Germany focus on specific questions or investigate the effectiveness of the job search, training, and other programs for a small number of participant groups and program types (e.g., Lechner et al., 2011; Kopf, 2013; Dengler, 2019; Goller et al., 2020). We contribute to this literature by evaluating the effectiveness of multiple job search and training programs for the German long-term unemployed. Importantly, we investigate the differential effectiveness of the ALMP for women and men separately. We provide fine granular heterogeneity in effects, individualized and on policy-relevant group levels, using a causal machine learning method for both genders. Moreover, we provide easy-to-implement rules to personalize the allocation of job search and training programs to the long-term unemployed.

---

[1] Lechner and Wiehler (2011) report similar estimates for Austria, but can explain a large part of the gender differences by pregnancies and parental leave.

For both ALMP for long-term unemployed and its' differential effects by gender, this is the first study that systematically estimates treatment effect heterogeneities of interest for policy and society. While estimating average effects for a population is well established in the microeconometric literature, systematically investigating treatment effect heterogeneities is a more challenging task. The recent, steadily growing literature on causal machine learning (CML), which combines predictive accuracy in statistical learning tools (for an overview, see Hastie et al., 2009) and causal principles that have been known for years, offers promising solutions to these challenges. In this study, we use the Modified Causal Forest (MCF)[2] proposed by Lechner (2018) and further analyzed by Lechner and Mareckova (2022). The MCF enables us to investigate multiple treatments and systematic effect heterogeneities on different levels of aggregation within one estimation approach.

Concerning optimal policy decisions, there is a growing theoretical literature on allocating treatments efficiently (see, e.g., Manski, 2004, 2007; Tetenov, 2012; Kitagawa and Tetenov, 2018; Athey and Wager, 2021). Such methods are increasingly used in empirical research, e.g., targeting coupon campaigns (Langen and Huber, 2023) or allocating vaccines (Kitagawa and Wang, 2023). In this study, we use both an optimal black-box allocation and an approach based on optimal policy trees (Zhou et al., 2022) to suggest easy-to-implement assignment rules for long-term unemployed to ALMP.

A special feature of our study is that we observe the universe of job search and training program participants among German means-tested benefit recipients. Having a broad range of administrative data for over 300,000 mainly long-term unemployed individuals, we investigate three different ALMP, namely *job-training*, *reducing impediments,* and (*private) placement services*, with participation starting in the first quarter of 2010. For those three ALMP, the unconfoundedness assumption used here for identification is credible as these administrative data are available, for which we provide additional evidence in a placebo test. The placebo exercise rejects the unconfoundedness hypothesis for a fourth program, *in-firm training*. This program is therefore excluded from the main analysis.

---

[2] While the estimation in this paper have been conducted with the Gauss version of the estimator, a Python version of it can be downloaded from PyPy.

When the programs we analyze were introduced in 2009, the German government attempted to create schemes that allow the job centers considerable leeway in the program's design to meet the individual needs of participants. In turn, participation should positively influence the employment perspectives of most participants.

We find that all investigated training and job search programs lead to positive employment effects on average and for most individuals. The average effect estimates are only partly higher for women than for men. Overall, effects emerge quickly and are long-lasting. While *placement services* is the program with the highest impact, we find substantial effect heterogeneity – especially for women. Generally, individuals with a worse labor market history benefit more than those with a better record. For women, the place of residence and local labor market conditions are decisive. Those located in regions with better local labor market conditions benefit substantially more from participating in job search and training programs than those women living in areas with worse labor market conditions. Regarding the allocation mechanism in place, we find a random allocation to perform equally well. While effect-based black-box allocation approaches lead to 14% higher effects for the reallocated subpopulation, even an easy and transparent rule leads to 6% higher effects.

The rest of the paper is structured as follows: Sections 2 and 3 introduce the institutional setting, the database used in the analysis, and some descriptive statistics. The methodology is described in Section 4, and the results of the empirical analysis are presented in Section 5. Allocation mechanisms are discussed in Section 6. Finally, Section 7 offers some concluding remarks.

## 2  Institutional setting

The German means-tested benefits (unemployment benefit II – UB II) are regulated in the system of basic income support called Social Code II (SC II). The official term of employment agencies in this system is "job centers," which we use in the following. Welfare recipients in this system are often long-term unemployed individuals running out of unemployment insurance benefits.[3] Although the pool of welfare recipients is rather heterogeneous, people with substantial employment impediments are frequently found among the unemployed receiving UB II. Both unemployed and employed individuals can receive UB II if their household income is below the poverty line, so they and their household members pass the means test. In 2010, the year of our observed treatment, UB II amounted to € 359 per month for a single adult (plus the costs for heating and accommodation).

---

[3] This is the other benefit type. The benefit level is 60% (67%) of the last net wage for childless adults (for parents). Unemployed persons receive this benefit up to one year if they are younger than 50 years (and up to two years for the older age groups).

ALMP play a major role in supporting unemployed welfare recipients in their integration into the labor market. The programs are supposed to help to increase their employability and labor market attachment. Therefore, caseworkers in job centers conduct profiling to identify strengths, like vocational skills or employment-relevant competencies, and employment impediments, like low education or individual life circumstances, of the unemployed. Based on the assigned profile[4], the caseworker and the unemployed agree on (personal) strategies to reach jointly defined short- and medium-term goals, such as employment or vocational training. ALMP assignment is often part of these strategies.

The ALMP of interest in this paper are specific subtypes of the "schemes for activation and integration" (SAI). SAI consist of different training programs within firms and in classrooms as well as placement services run by private providers. The job centers have considerable leeway in the program's design. This allows them to adapt the program to the individual skills and situations of the participants (see Goller et al., 2020; Harrer et al., 2020).

We particularly focus on the following four SAI subtypes[5]: (1) guiding into apprenticeship and work (in the following "*job-training*" or *JT*), (2) determining, reducing, and removing employment impediments (in the following "*reducing impediments,*" or *RIM*), (3) placement into contributory employment (in the following "*placement services,*" or *PS*), and (4) *in-firm training* (*IFT*). *JT* and *RIM* both take place at private training providers; *PS* is conducted by private placement services, while companies organize *IFT* as a type of internship.

During *job-training*, participants learn to choose suitable job offers and to write application letters and CVs. This is supposed to improve their job search effectiveness. *Reducing impediments* focuses on the participants' skills and employability. *RIM* aims at overcoming the participants' employment impediments by increasing participants' knowledge about specific occupational fields, for example. *Placement services,* in contrast, aim at finding work or vocational training for participants. *IFT* allows participants to get accustomed to regular work schedules and the employment situation in the company hosting *IFT*. The duration of these programs is relatively short. 99.0% of *IFT* inflows between January and March 2010 had a program duration of less than one month.[6] For *JT* and *RIM* inflows, the shares of programs lasting less than three months were 92.9% and 94.5%, respectively. With 53.2%, the respective shares were smaller for *PS* inflows between January and March 2010.

---

[4] Such as the market profile for people who are highly attached to the labor market or the development profile for people with a low labor market attachment.

[5] We did not include other subtypes for different reasons. This was because of too few observations, a very selective targeting, a quite heterogeneous design with insufficient information on the programs' content, i.e., impeding proper interpretation of our results.

As already mentioned, a placebo exercise rejected the unconfoundedness hypothesis for *IFT*, so we focused our main analyses on the SAI subtypes of *JT*, *RIM,* and *PS*, which are programs characterized by relatively homogeneous treatments and well-defined program content. The distinctive subtypes differing in their aims allow us to answer whether participants might have improved their labor market chances if assigned to another treatment. Moreover, the three subtypes are important in practice, providing us with sufficiently high numbers of observations to use the elaborate econometric methods of machine learning and group analyses.

## 3 Data

Our rich administrative dataset stems from the Statistics Department of the German Federal Employment Agency and contains information on (registered) jobseekers and benefit recipients. We were able to use a rich set of observable characteristics relevant to welfare recipients' labor market integration as we included information at the individual, the household, the district, and the job center level.[7]

In detail, we included sociodemographic characteristics, a large set of variables on the labor market history of our sample members, information on the last job and the labor market status in December 2004, i.e., before the introduction of the SC II. Among the covariates at the individual level, we included variables such as age, gender, children living in the same household, the last occupation, and work experience. Older age, care responsibilities, and work experience made long ago decrease the probability of leaving unemployment and welfare receipt (Hohmeyer and Lietzmann, 2020) or at least slow down the transition from welfare receipt into self-sufficient employment (Achatz and Trappmann, 2011; Beste and Trappmann, 2016). Including rich information on the labor market biographies (e.g., not only on work experience but also on unemployment and ALMP program experience), we indirectly controlled for unobservable characteristics such as motivation or personality traits (Caliendo et al., 2017).

---

[6] Source: DataWareHouse of the Statistics Department of the German Federal Employment Agency.

[7] The full list of the covariates, including descriptive statistics, can be found in Online Appendix O-D.

At the household level, we considered the income and composition of the household. We controlled for the number and age of children in the household as (high numbers of) young children diminish the chances to exit unemployment and welfare receipt (Hohmeyer and Lietzmann, 2020). We further differentiate by gender in our analyses because care responsibilities due to such compositional situations are more likely to negatively affect women's labor market prospects than men's (Achatz and Trappmann, 2011). Further, (potential) welfare receipt might affect household composition in SC II (due to the amount of received benefits). Possible composition changes (e.g., divorce, cohabitation, or birth of children) may lead to changes in the household's risk of welfare receipt (Blank, 1989). We also included information on the partner if living in the same household. The partner's work experience and education determine their employment prospects, affecting the household's chances of leaving UB II. Lastly, we included district-level labor market indicators, such as the unemployment rate, and information at the job center level, such as the client-staff ratio in the job centers. We did so because the local situations in job centers and labor markets are likely to influence welfare recipients' labor market prospects (e.g., found by Carpentier et al. (2014) in observing social assistance exit rates in Belgium).

Our sample consists of individuals who were unemployed and received UB II at the end of 2009. We further modified this dataset to get our final sample. The three treatment groups used in the main analyses consist of the population of unemployed welfare recipients starting *JT*, *RIM*, or *PS* in the first quarter of 2010. Moreover, a group of *non-participants* (*NP*) representing a 20% random sample of the stock of unemployed UB II recipients at the end of 2009 who did not enter any SAI program during the following three months. If individuals participated in several of our observed SAI subtypes, the very first of these subtypes determines to which of the three treatment groups the individuals belong.

*Table 1: Descriptive statistics – Outcome and selected covariates*

|  | Men | | Women | |
|---|---|---|---|---|
| Variable | Participants | Non-participants | Participants | Non-participants |
| Cumulated days in regular employment in the 3 years after treatment (Outcome) | 250 (325) | 181 (291) | 188 (300) | 137 (261) |
| Personal characteristics | | | | |
| Age at sampling date (in years) | 38.3 (8.52) | 39.7 (8.57) | 38.9 (8.38) | 39.9 (8.41) |
| Days since last employment | 1,478 (1,598) | 1,822 (1,774) | 1,949 (2,128) | 2,115 (2,287) |
| Days in regular employment in the previous 5 years | 293 (401) | 224 (358) | 178 (343) | 138 (298) |
| Days in welfare receipt in the last year | 297 (112) | 318 (91) | 317 (98) | 332 (79) |
| Receiving income from dependent employment (yes=1) | 0.17 | 0.18 | 0.22 | 0.27 |
| Region (west=0, east=1) | 0.25 | 0.35 | 0.26 | 0.33 |
| Foreigner (yes=1) | 0.23 | 0.20 | 0.20 | 0.22 |
| Job center characteristics | | | | |
| Job center district - client-staff ratio | 158 (28) | 162 (26) | 159 (28) | 162 (27) |
| Sanction intensity due to violations of duties (in %) | 0.54 (0.25) | 0.51 (0.24) | 0.54 (0.25) | 0.52 (0.24) |
| Sanction intensity due to failure in reporting (in %) | 0.71 (0.25) | 0.71 (0.25) | 0.71 (0.26) | 0.71 (0.26) |
| District-level characteristics | | | | |
| District unemployment rate (in %) | 10.4 (3.3) | 11.0 (3.5) | 10.3 (3.3) | 10.8 (3.6) |
| District unemployment rate of welfare recipients (in %) | 7.4 (3.0) | 7.9 (3.3) | 7.4 (3.0) | 7.7 (3.3) |
| N | 28,425 | 136,691 | 20,741 | 116,769 |

Note. – Means of the covariates. Standard deviations are in parentheses. The treated group in this table contains all individuals from our initial three treatment groups (*JT*, *RIM*, and *PS*). We computed the mean of covariates over all treatment groups here.

We only included unemployed welfare recipients aged 25 to 54 due to the different ALMP assignment rules the Federal Employment Agency has for younger and older welfare recipients. In our observation window, special rules for individuals younger than 25 lead to more intense activation than for older welfare recipients. Moreover, by excluding very young welfare recipients, we also make sure to get (un)employment biographies more complete, thus being able to indirectly control for unobservable characteristics that are highly related to the employment biographies. As we focus on *JT*, *RIM*, and *PS* in this study, we excluded welfare recipients participating in one of the other SAI subtypes during our treatment window (January to March 2010).

We excluded individuals who found contributory employment or left welfare receipt between the sampling date and their (hypothetical) program start. Measuring outcomes for the treated individuals from their program start onward is straightforward. To compare participants with non-participants, we would have liked to measure the outcomes for the latter in the same way, but no program start is available for them. This was resolved by assigning a hypothetical program start to each of the non-participants. It was randomly drawn from the distribution of actual program starts among participants (similar to, e.g., Gerfin and Lechner, 2002; Sianesi, 2004; and Goller et al., 2020). Finally, we deleted observations from our sample due to missing values in the covariates. All in all, this leads to 302,626 observed individuals (see Table 1).

In Table 1, we present some selected covariates and distinguish between non-participants and a combined group of treated sample members. On average, participants show more beneficial characteristics than non-participants, i.e., participants experienced less cumulated days in welfare receipt in the last year but more days in regular employment in the previous five years.

## 4 Econometrics

### 4.1 Notation and framework

To describe our multiple treatment model under conditional independence (Imbens, 2000; Lechner, 2001), we use Rubin's (1974) potential outcome framework. Participation in one of the programs is indicated with $D_i$ as the (multiple) treatment variable, while $D_i = 0$ indicates non-participation of the individual $i$ ($i = 1, .., N$) and $D_i > 0$ participation in one of the three job search and training programs. Let $Y_i^d \coloneqq Y_i(D_i = d)$ denote the potential outcome if individual $i$ receives treatment $d \in \{0,1,2,3\}$.[8] For each individual, we observe the particular potential outcome related to the treatment status to which the individual is assigned; the others remain counterfactual: $Y_i = \sum_{d=0}^{3} \mathbf{1}(D_i = d) Y_i^d$. Further, for each individual, we observe the variables $X_i \in (\tilde{X}_i, Z_i)$. While $\tilde{X}_i$ represents those variables needed to account for confounding, $Z_i$ contains those variables in which we are interested in the heterogeneity analysis.[9]

There are three estimands of interest on different levels of aggregation:

$$IATE(m, l; x, \Delta) = E(Y^m - Y^l | X_i = x, D_i \in \Delta),$$

$$GATE(m, l; z, \Delta) = E(Y^m - Y^l | Z_i = z, D_i \in \Delta),$$

$$ATE(m, l; \Delta) = E(Y^m - Y^l | D_i \in \Delta).$$

The Average Treatment Effects (ATEs) represent the population average effects on the highest level of aggregation for treatment status $m$ compared to treatment status $l$ belonging to treatment groups $\Delta$, where $\Delta$ denotes all treatments of interest. Please note that if $\Delta$ relates to the population $D = m$ we obtain the Average Treatment Effect on the Treated (ATET). On the contrary, the estimand on the lowest aggregation level is the Individualized Average Treatment Effect (IATE), i.e., conditional on characteristics $x$. An estimand on the intermediate aggregation level, which is of main interest for policy analysis, is the Group Average Treatment Effect (GATE) according to the heterogeneity variables $Z_i$. Both special cases of the Conditional Average Treatment Effects (CATEs) last mentioned, the GATEs and IATEs, are helpful to detect heterogeneities, which are otherwise "hidden" in the homogenous ATE estimate. Worth noting is the relationship between those three estimands. Averaging the IATEs by the groups $Z_i = z$ results in the GATEs. Averaging over the GATEs then leads to the ATEs.

## 4.2 Identification

As mentioned, only one of the potential outcomes is observable, since precisely one of the four treatment statuses can be realized, the others remain counterfactual. In the literature, this is referred to as the 'fundamental problem of causal inference' (Holland, 1986). A credible identification strategy for estimating causal effects is crucial to 'solving' this. We rely on a

---

[8] We use the convention that (usually) capital letters denote random variable, while small letters denote some fixed value of these random variables.

[9] In principle, $\tilde{X}_i$ and $Z_i$ might contain distinct variables or overlap, partly or completely. In this work $Z_i$ is a subset of $\tilde{X}_i$ selected ad hoc. The heterogeneous treatment effects that are based on this selection of variables are of considerable interest for policy makers, society, and academia.

selection-on-observable approach and need to impose some assumptions to identify the estimands of interest in our multiple treatment setting (see Imbens, 2000; Lechner, 2001):

CIA: $Y^d \perp D | \tilde{X}_i = \tilde{x}$,

CS: $0 < P(D_i = d | \tilde{X}_i = \tilde{x}) < 1$,

SUTVA: $Y = \underline{1}(D_i = d) Y^d$,

The Conditional Independence Assumption (CIA) states that all the potential outcomes are independent of the treatment assignment, conditional on the observed confounders. This implies that no other characteristics are jointly related to the potential outcomes and treatment. Common Support (CS) is ensured if every treatment status might be observed for all realizations of covariates. The Stable Unit Treatment Value Assumption (SUTVA) requires no spill-over effects across the treatment groups.

As discussed in Section 3, the available covariates capture a wide range of potential confounders. Most of the quantifiable information that the caseworker responsible for the assignment to the programs can see is contained in our data set. Among those are the most critical confounders, as identified by other evaluation studies (Heckman et al., 1998; Lechner and Wunsch, 2013). In addition, we include more characteristics related to the individuals' labor market history in the last five years since means-tested benefit recipients mostly consist of people who did not work for various years. A strength of our data set is that we can control for an assessment variable defining the profile of the unemployed, which results from a usually unobservable profiling process, including talks between the caseworker and the unemployed. Therefore, the CIA in this work is arguably credible with this rich administrative data set. While this assumption is untestable, we provide a placebo study to strengthen the argument.[10] Since the observed programs are rather small compared to the labor force, there should not be any spill-over effects rendering the SUTVA incredible. Exogeneity is given as we measure all covariates at our sampling date (12/31/2009) before they are assigned to any treatment status.

## 4.3 Method

Recently, many new estimators were proposed, combining the predictive power of machine learning tools and the causal structure known from classical microeconometric literature (e.g., Athey, 2018; Athey and Imbens, 2019). Those methods, branded as CML, turn out to be especially useful for estimating treatment effects beyond the average effects.

---

[10] The placebo study does provide evidence for the CIA to hold for the three treatments investigated. It also confirms our choice to not include another, fourth treatment *IFT*, for which the assignment mechanism is probably driven by some external factors, which are not observable for us.

In our empirical analysis, the MCF (Lechner and Mareckova, 2022), a well-performing estimator for multiple treatments, which provides estimates on the various levels of aggregation and inference for those parameters, is used. Simulation-based evidence (e.g., Knaus et al., 2021) finds a general observation that forest-based CML estimators perform especially well in many situations (Wager and Athey, 2018; Athey et al., 2019).

The foundation of forest-based methods was laid by the Random Forest introduced by Breiman (2001) as an ensemble of many regression trees. The idea of a regression tree is to recursively split the space of covariates into non-overlapping areas by minimizing the mean squared error (MSE) of the outcome prediction until some stopping criteria are reached. The resulting structure is reminiscent of a rotated tree, as one observes the trunk with all the observations in the beginning, split up into finer branches the further one goes down. The final predictions result from the averages of the outcomes falling into the same end nodes, called leaves. The combination of many randomly constructed trees gives the final prediction of the random forest. Athey and Imbens (2016) developed a causal tree to accommodate this prediction tool in the causal framework. Many of those causal trees can be combined into a Causal Forest in different forms (Wager and Athey, 2018; Athey et al., 2019).

Lechner (2018) and Lechner and Mareckova (2022) further develop this idea by improving the splitting rule for the individual trees and providing weight-based inference. This new estimator, called MCF, is especially well-suited for this study. It enables the estimation of heterogeneous effects in a multiple-treatment setting for various levels of aggregations. With an approach for unified inference for the highest (ATE) and lowest level (IATE) of aggregation as well as the intermediate level for variables of policy interest (GATE), this estimator is computationally attractive and well-fitted for the empirical challenges in analyzing active labor market policies. The interested reader is referred to Lechner and Mareckova (2022) for technical details and Online Appendix O-C for details on the implementation.

## 5 Results

We report and discuss the main results, while additional results not shown can be found in Online Appendix O-A. First, the usual population average effects are discussed, as is also done in most previous empirical research in ALMP evaluation. This enables us to evaluate the overall effectiveness, in terms of additional days in regular employment, of the three investigated training and job search programs, compared to each other and non-participation in any program. Second, we investigate more fine granular effects on the policy-relevant group average and individualized levels.

## 5.1 Average effects

Table 2 presents the ATEs for the three different programs (*job-training*, or *JT*; *reducing impediments*, or *RIM*; *placement services*, or *PS*) and *non-participation* (*NP*) against each other and ATETs for the respective participants' groups. The outcome is cumulated days in regular employment in the three years after the treatment start.

*Table 2: Average Treatment Effects*

|     | Men | | | | | Women | | | | |
| --- | --- | --- | --- | --- | --- | --- | --- | --- | --- | --- |
|     | NP | JT | RIM | PS | ATET | NP | JT | RIM | PS | ATET |
| Cumulated Days in regular Employment in 36 months after start of treatment (outcome) | | | | | | | | | | |
| NP  | 181.1 | | | | | 137.6 | | | | |
|     | (2.1) | | | | | (2.2) | | | | |
| JT  | 36.3*** | 217.4 | | | 35.0*** | 25.7*** | 163.3 | | | 23.6*** |
|     | (5.4) | (4.9) | | | (6.2) | (5.6) | (5.1) | | | (5.9) |
| RIM | 34.6*** | -1.7 | 215.7 | | 34.1*** | 36.8*** | 11.2 | 174.4 | | 34.7*** |
|     | (6.3) | (7.8) | (6.2) | | (6.8) | (6.7) | (8.3) | (6.4) | | (6.9) |
| PS  | 45.1*** | 8.8 | 10.5 | 226.2 | 46.7*** | 54.1*** | 28.5*** | 17.3 | 191.7 | 64.1*** |
|     | (7.8) | (9.0) | (9.6) | (7.0) | (7.6) | (8.7) | (10.0) | (10.7) | (8.4) | (9.1) |
| N   | 136,691 | 12,329 | 8,721 | 7,375 | | 116,769 | 9,350 | 6,678 | 4,713 | |

Note. – Outcomes are measured in days in regular employment after starting the treatment. ATET relative to no treatment only; Potential outcomes are on the main diagonals. Standard errors are in parentheses. *** indicates that the p-value of a two-sided significance test is below 1%. The programs are labeled as NP: non-participation, JT: job-training, RIM: reducing impediments, and PS: placement services.

We find all programs to be effective compared to non-participation for men and women, with *PS* as the most effective for men (45.1 days) and women (54.1 days more in regular employment). For men, the effects are not significantly different when the different programs are compared. For example, *RIM* is 1.7 days less effective than *JT*, but with a standard error of 7.8, hence the difference is statistically insignificant. *PS* is preferable to *JT* for women, resulting in 28.5 days more in regular employment if allocated to *PS* compared to *JT*. The potential outcomes in the respective treatments are documented on the main diagonal of Table 2. In general, they are higher for men than for women, which implies, that the estimated effects are relative to their baseline outcome higher for women compared to men.

First insights into the efficiency of the existing allocation mechanism can be obtained by investigating how the effects from Table 2 differ for the different subpopulations of participants in the three programs. Caseworkers effectively allocate individuals to programs if the group of selected individuals to a specific treatment benefit more from the treatment than the non-selected individuals. In other words, if the ATET is larger than the ATE (or the average treatment effects of those groups selected to none or a different training program), the treated population is well chosen. For *PS* compared to non-participation, the caseworkers' selection mechanism tends to be effective, i.e., the ATET is larger than the ATE. Compared to non-participation, the selection mechanisms tend to be rather ineffective for *RIM* and *JT*. While this cannot tell us anything about the total effect, it indicates a substantial potential to allocate participants to the specific programs in a well-informed way, which is the topic of Section 6.

*Figure 1: Evolution of ATE over time*

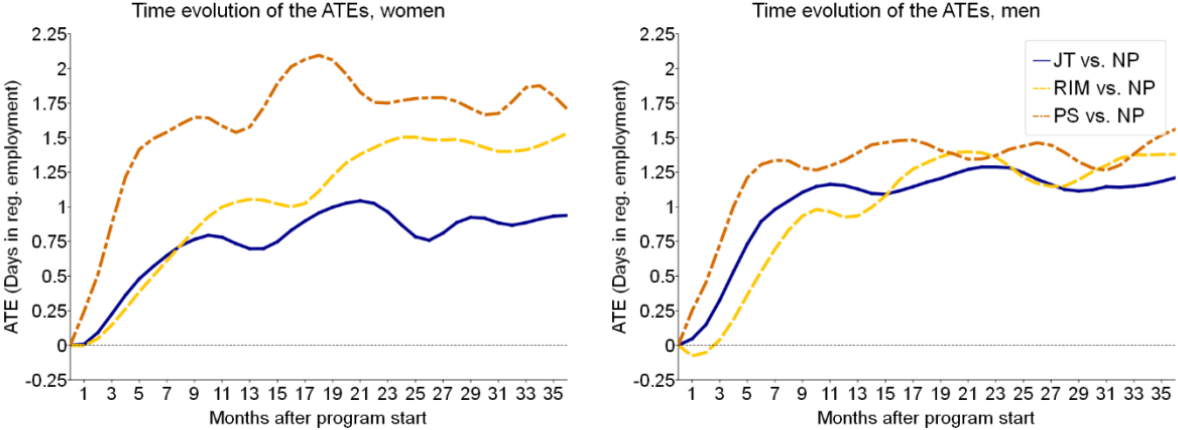

Note. – Outcome is the cumulative days in regular employment for each month after the start of participation in the respective program. After month 4, all estimates significantly differ from zero at conventional levels. Exemplary standard errors for month 10 (women: 0.19, 0.23, 0.31; men: 0.19, 0.23, 0.28), month 20 (women: 0.24, 0.28, 0.37; men: 0.23, 0.27, 0.33), and month 30 (women: 0.25, 0.30, 0.39; men: 0.24, 0.28, 0.33) for JT vs. NP, RIM vs. NP, and PS vs. NP, respectively.

To complete the picture, Figure 1 presents how the effects of participation in specific programs versus non-participation evolve over time. First, we do not observe severe lock-in effects for any of the programs. This is expected as the data mainly covers long-term unemployed individuals, for whom it is hard to find a job in general and who often show higher program participation effects. Second, the programs' durations are relatively short, i.e., program participation keeps welfare recipients from job search only for a short time. Lastly, *PS* is always the most beneficial for women, with a quick realization period in the first year and a relatively constant benefit of about 1.6 - 2.0 more days in employment per month for participants after one year.

This finding is not surprising as *PS* is designed to push participants into the labor market. *RIM* steadily increases employment until 36 months after participation. Such training programs often unfold their effectiveness in the medium to long run (see Lorentzen and Dahl (2005) for Norway or the meta-analyses of Card et al. (2010, 2018)). The effect of *JT* increases to about 0.8 additional days in employment after nine months and remains at this low level. Further, the participation effects of this program type often stay relatively constant in the medium to long run.

For men, the picture is different. While *PS* is the most beneficial program in the first months, the effects of *JT* and *RIM* catch up and remain close to each other. Especially the effect of *PS* is lower for men throughout, compared to women (see similar findings in, e.g., Bergemann and van den Berg, 2008; Dengler, 2019), leading to the benefit of about 1.1 - 1.4 more days in employment per month for participating in any program for men in the long run.

The evidence presented for the population level documents the overall effectiveness of the programs as well as the potential for better allocation of participants to the programs. Next, we investigate lower levels of aggregation by starting with an intermediate level for general or specific characteristics of political interest. After that, we continue with effects on the (almost) individual level.

## 5.2 Group effects

The analysis on the GATE level is of particular policy interest and an excellent way to investigate treatment effects on an intermediate level systematically. Using all available potential variables for a heterogeneity analysis and reporting significant results would be data mining and surely the wrong way. We, therefore, pre-specified a shortlist of variables of particular interest for policymakers, society, and academia in Table 3.

Table 3 shows Wald test p-values for men and women, indicating whether there are differential effects for groups of individuals for the specific variable. A p-value below 10 / 5 / 1% rejects the null hypothesis of non-differential effects for the respective groups of individuals. While this is not a formal test for effect heterogeneity, it can indicate differential effects. Significant differential effects in Table 3 can be found mainly for women, specifically for *JT* and *PS* vs. *NP*. Variables indicated with "§" are discussed in the following, while results for the other variables are shown in Online Appendix O-A.

*Table 3: Short List - Wald tests for heterogeneity (p-values)*

| Variable | Women | | | Men | | |
|---|---|---|---|---|---|---|
| | JT vs. NP | RIM vs. NP | PS vs. NP | JT vs. NP | RIM vs. NP | PS vs. NP |
| Personal characteristics | | | | | | |
| Age§ | 93 | 46 | 13 | 98 | 86 | 93 |
| Family status | 3 | 73 | 1 | 16 | 58 | 59 |
| Educational achievement | 14 | 24 | 6 | 2 | 61 | 95 |
| Days in reg. empl. in the prev. 5 yrs§ | 99 | 83 | 70 | 94 | 85 | 65 |
| District-level characteristics | | | | | | |
| Region (East vs. West)§ | 1 | 42 | <1 | 38 | 66 | 93 |
| District unemploym. rate§ | <1 | 67 | <1 | 31 | 99 | 99 |
| Job center characteristics | | | | | | |
| Client-staff ratio | 8 | 63 | 5 | 79 | 99 | 99 |
| Sanction intensity – viol. of duties§ | <1 | 96 | 2 | 35 | 99 | 99 |

Note. – P-values in % from Wald tests for heterogeneity. Variables indicated with § are discussed in the following; results for the other variables are shown in Online Appendix O-A.

Due to historical reasons in many fields of the economy, there is still some discrepancy between the former German Democratic Republic region and the western part of Germany. It is, therefore, interesting to analyze if training and job search programs today are equally, more, or less effective in either of those regions. We complement this with regional economic conditions since the economic conditions in East and West Germany are still very different, e.g., higher rates of unemployment and fewer vacancies in the East. Further, especially important for the labor market authorities are the labor market history of the individual and the age because those are well-observable characteristics and might indicate the general potential of the unemployed. The third party involved are the job centers. Here, we are interested in the influence of the sanction intensity, i.e., how many sanctions are imposed in the specific job center as an indication of the toughness of the caseworkers. All figures presented show differences of GATEs to the respective ATE, for women and men separately, to investigate how those effects differ compared to the average and with regard to gender.[11]

*Figure 2: Difference of GATEs to ATE in West / East Germany*

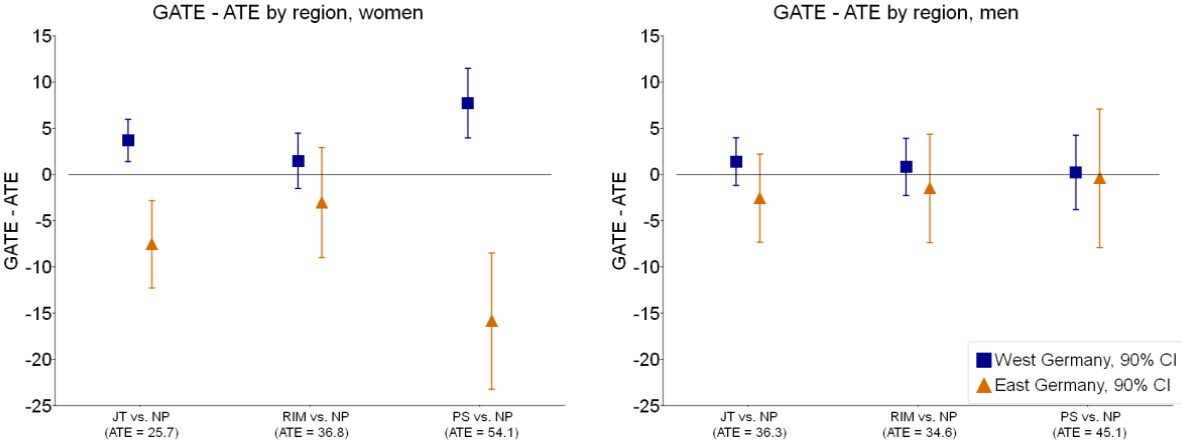

Note. – 90% confidence intervals (CI). ATE estimates are in parentheses at the bottom of the graphs. The outcome is the cumulative days in regular employment in 36 months after the start of treatment.

Figure 2 shows the treatment effects of the programs against non-participation by regions. We find the effects for the group of individuals in the western part of Germany to be higher than those in the eastern part in every comparison. Therefore, programs, in general, seem to work better in West Germany compared to East Germany. Bernhard and Wolff (2008) found the same pattern for contracting out to private placement providers (program replaced by *PS*). However, our study finds significantly differential effects for women in *JT* and *PS* compared to non-participation. At the same time, the GATEs are all close to the population average effects for men.

To investigate if this is driven directly by the local economic conditions, we analyze the GATEs associated with the district unemployment rate (Lechner and Wunsch, 2009). Indeed, in general, treatment effects are higher for those female participants in regions with a lower unemployment rate, as shown in Figure 3 exemplarily for *PS* vs. *NP*. As *PS* needs vacancies to unfold its effectiveness and lower unemployment rates might be correlated with higher amounts of available jobs, this finding is clear-cut. For men, the results of Figure 3 align with the findings in Figure 2, as the GATEs fluctuate around the ATE. The same pattern can be observed for GATEs associated with welfare recipients' unemployment rate.

---

11 To obtain the "pure" GATE, the ATE provided in each figure has to be added to the point estimate of the respective group. Since the vast majority of the GATEs are significantly different from zero, we opted for presenting the GATE-ATE to focus explicitly on within group differential effects.

*Figure 3: Difference of GATEs to ATE by the district unemployment rate, PS vs. NP*

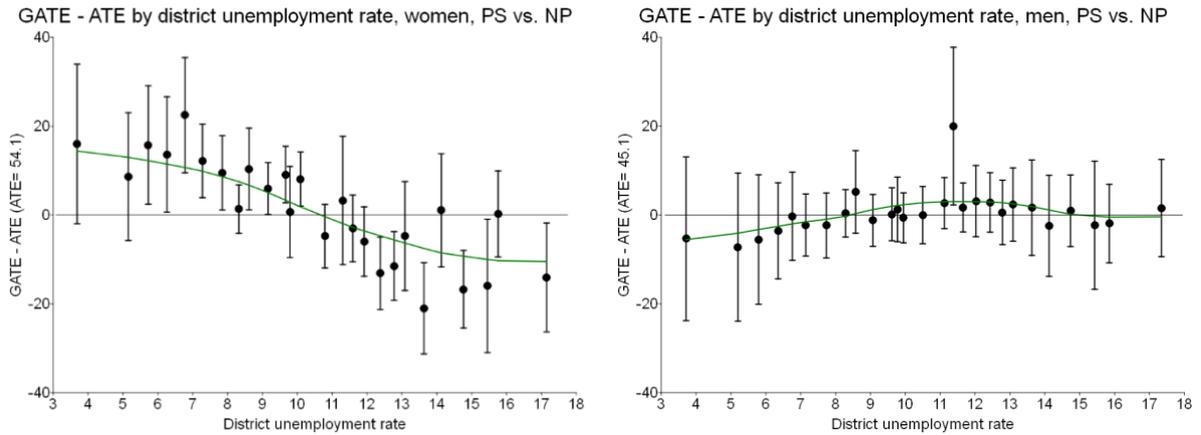

Note. – 90% confidence intervals. ATE estimates are in parentheses on the left-hand side of the graphs. The outcome is cumulative days in regular employment 36 months after the start of treatment. The green line represents the (kernel-)smoothed GATE estimates and is for illustration purposes only.

For previous labor market success, in the form of days in regular employment in the previous five years, results can be found in Online Appendix O-A. We do not see any clear pattern. In addition, confidence intervals become increasingly large due to fewer observations with many days in regular employment in the last five years. Thus, we refrain from further interpretation.

*Figure 4: Difference of GATEs to ATE by age, RIM vs. JT*

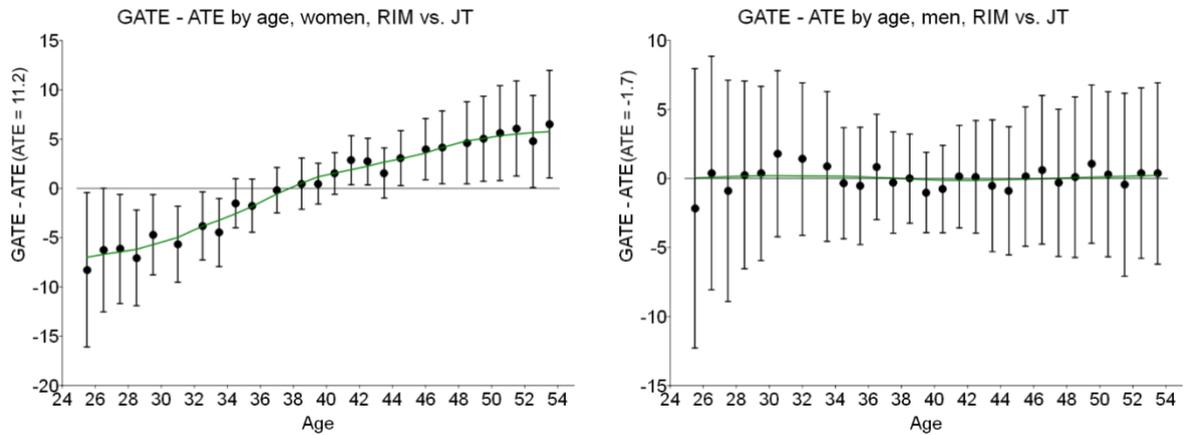

Note. – 90% confidence intervals. ATE estimates are in parentheses on the left-hand side of the graphs. The outcome is cumulative days in regular employment 36 months after the start of treatment. The green line represents the (kernel-)smoothed GATE estimates and is for illustration purposes only.

Relevant for practitioners is whether younger and older people should be sent to the same or different ALMP. Figure 4 shows differential effects with individuals grouped by age for sending them to *RIM* compared to *JT*. While it does not seem to matter for men, with all GATEs around the ATE, this does matter for women. Younger women should rather be allocated to the *JT* or *NP,* older women benefit more from the *RIM* program. Instead, *PS* seems to be more beneficial for older men than *NP*, *JT,* or *RIM*, while the latter two are more beneficial for younger men; results for this and the other comparisons can be found in Online Appendix O-A. That *JT* is more beneficial for younger participants is in line with Caliendo and Schmidl (2016), who found positive effects for young people participating in (solely) classroom-based training programs (to which *JT* is the most similar of our observed programs).

*Figure 5: Difference of GATEs to ATE by job center sanction intensity, JT vs. NP*

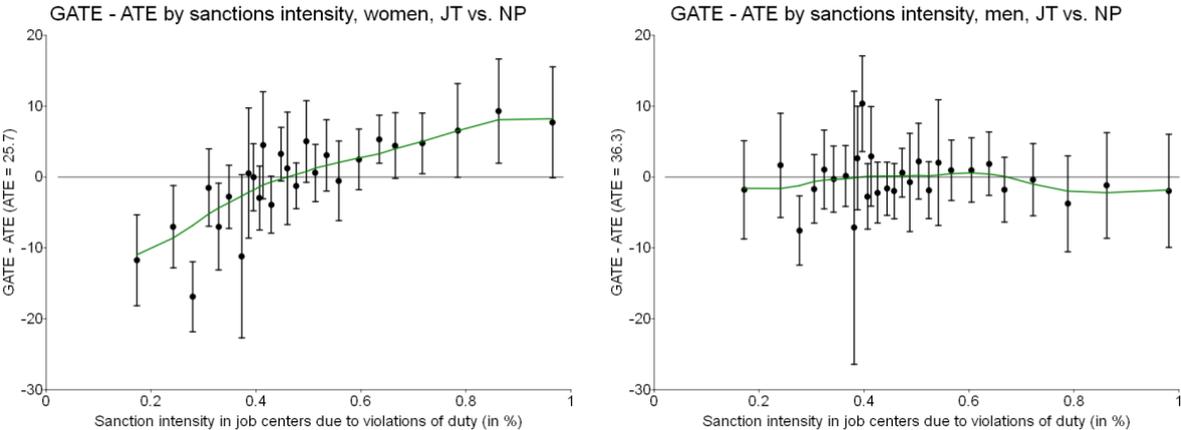

Note. – Job centers' sanction intensity due to violations of duty. 90% confidence intervals. ATE estimates are in parentheses on the left-hand side of the graphs. One outlier group is omitted for the sake of visibility. The complete graph can be found in Online Appendix O-A. The outcome is cumulative days in regular employment 36 months after the start of treatment. The green line represents the (kernel-)smoothed GATE estimates and is for illustration purposes only.

For job centers, imposing sanctions is a controversial tool to encourage means-tested benefits recipients to search for jobs actively, participate in training programs, etc. From an academic point of view, it is interesting to investigate if those unemployed, supported by job centers imposing more sanctions, benefit more or less from participation. Figure 5 provides results for being allocated to *JT,* compared to not being allocated to one of the training programs, associated with the job centers' sanction intensity. We find that the effects for women associated with job centers with a higher sanctions intensity are above the average effect. In comparison, those associated with a job center imposing fewer sanctions benefit below average. Again, for unemployed men, there are no differential effects. The same pattern is found for comparing the other treatments to non-participation.

## 5.3 Individualized average effects

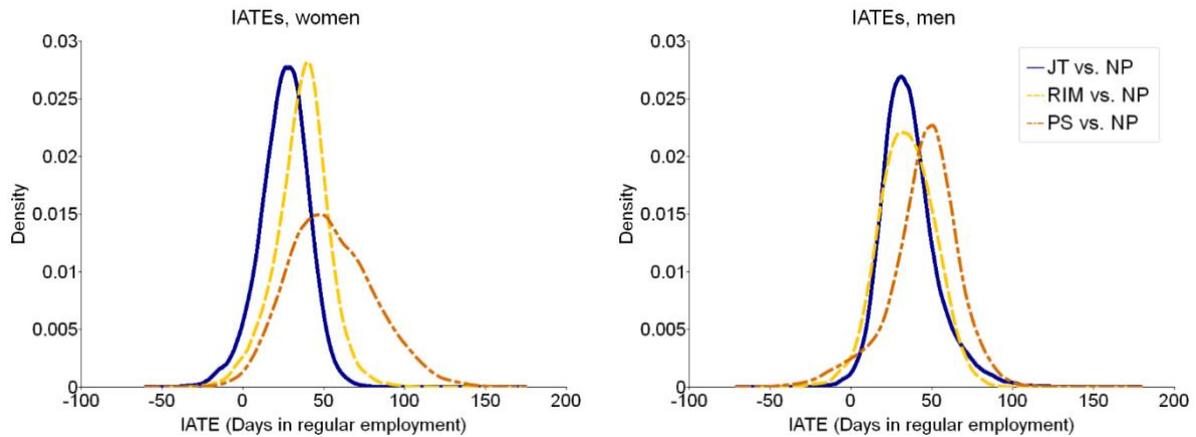

*Figure 6: Distribution of estimated IATEs*

Note. – IATE density plots. The outcome is cumulative days in regular employment in 36 months after the start of treatment.

*Table 4: IATEs, descriptions*

|  | Women | | | | Men | | | |
|---|---|---|---|---|---|---|---|---|
|  | Share >0 | Share >0 (**) | Std | SE (aver.) | Share >0 | Share >0 (**) | Std | SE (aver.) |
| JT vs. NP | 94.5 % | 44.4 % | 15.14 | 16.18 | 99.2 % | 59.1 % | 17.39 | 18.69 |
| RIM vs. NP | 97.7 % | 55.2 % | 16.58 | 19.48 | 97.6 % | 49.1 % | 17.54 | 21.44 |
| PS vs. NP | 99.0 % | 63.8 % | 26.85 | 24.53 | 95.7 % | 57.6 % | 21.98 | 24.98 |

Note. – Proportion (Share) of positive IATEs. A proportion significantly larger than zero is indicated with ** on the 5% level. Std stands for the standard deviation of the respective distribution, and SE (aver.) is the standard error averaged over all IATEs.

On the lowest level of aggregation, Figure 6 presents the distributions of estimated IATEs for participating in one of the treatments against non-participation. The first observation is that most individuals realize positive effects for all labor market programs. Table 4 documents this with shares of 94.5 - 99.2% of individuals having a positive impact from participating in either program. This is manifested in the substantial proportions of individuals with significant positive effects. Another conclusion from Figure 6 is that *PS* leads to the largest gains in cumulated days in regular employment on average, as discussed in the previous subsection, and for a substantial share of individuals.

Comparing the distributions of men with those of women, it is especially noteworthy that the distribution of *PS* for women is wider than for men, as apparent from the standard deviations (Std) shown in Table 4. This might be attributed to some degree of estimation error. However,

Table 4 shows the highest average standard errors for both men and women for *PS* vs. *NP*–this cannot explain the wider distribution for women compared to men. The remaining explanation is that it points to considerable effect heterogeneity in this program and, thus, high potential for a more beneficial allocation of training programs.

To describe those populations of individuals benefiting most and least from ALMP participation, we show the dependence of the effects on characteristics by k-means++ clustering (compare Arthur and Vassilvitskii, 2007). For most and least affected individuals, clusters are formed by jointly using the IATEs of participation in one of the programs relative to non-participation. The approach is detailed in Appendix A, and the results are shown in Table A.1. Summarizing the analysis, we can draw three general indications. First, the most and least benefiting populations are different for men and women. Second, women residing in areas with better local labor market conditions are in the group that benefits most. Third, good a priori risks are in the group that benefits least from participating in the labor market programs.

## 5.4 Placebo analysis

The credibility of empirical results depends on the validity of the underlying assumptions. While the unconfoundedness assumption for the training and job search programs investigated with such extensive administrative data is very likely to be fulfilled in our particular setting, it is not directly testable. An indirect approach to assess the validity of the CIA is to conduct a placebo analysis, as described by Imbens and Wooldridge (2009) for example.

To be more concrete, we take those individuals of our original sample who were also unemployed two years before the treatment is allocated, i.e., by the end of 2007, and measure the outcome as cumulative days in regular employment in one year after this pseudo-treatment date.[12] For this, we constructed a new database with most of the covariates used in the main analysis dated back to the end of 2007. Participation in the job search or training programs in 2010 should not have an influence on the outcome measured prior to this. Therefore, if we observe all confounding characteristics, the pseudo-treatment effect should be zero. Not rejecting this placebo null hypothesis does not imply that the CIA is valid but gives some evidence that the CIA is plausible in this case. If we reject this test, there might be some unobserved confounding.

---

[12] We cannot use the cumulative days in employment for three years, as this is partly in our true treatment phase. Moreover, we cannot go further into the past, since we do not observe many of the covariates that are related to the benefit system otherwise. The set of covariates is slightly smaller now, as we cannot construct all the covariates back two years before the treatment. Still, the most crucial variables are captured in our pseudo-analysis database. If we do not reject the null hypothesis with a smaller set of covariates, having more is likely to make the unconfoundedness assumption even more plausible.

*Table 5: Placebo effects for future programs on cumulative days in regular employment*

|      | Men |     |     |     | Women |     |     |     |
|------|-----|-----|-----|-----|-------|-----|-----|-----|
|      | NP  | JT  | RIM | PS  | NP    | JT  | RIM | PS  |
| JT   | 1.1 |     |     |     | 1.1   |     |     |     |
|      | (1.3) |   |     |     | (1.1) |     |     |     |
| RIM  | 0.8 | -0.3 |    |     | 0.9   | -0.2 |    |     |
|      | (1.4) | (1.7) |  |     | (1.1) | (1.5) |  |     |
| PS   | 2.1 | 1.1 | 1.3 |    | 2.3   | 1.1 | 1.4 (1.9) | |
|      | (1.8) | (2.1) | (2.2) | | (1.6) | (1.8) |  |     |
| IFT  | 14.9*** | 13.8*** | 14.1*** | 12.8*** | 9.0*** | 7.9*** | 8.2*** | 6.8*** |
|      | (1.6) | (1.9) | (2.0) | (2.3) | (1.5) | (1.7) | (1.8) | (2.1) |

Note. – Outcome is cumulated days in regular employment in one year after pseudo-treatment. Standard errors are in parentheses. *** indicates that the p-value of a two-sided significance test is below 1%. The programs are labeled as NP: *non-participation*, JT: *job-training*, RIM: *reducing impediments*, PS: *placement services*, and IFT: i*n-firm training*.

Table 5 presents the results for all four initial programs and *non-participation,* investigated in the placebo analysis. Besides Treatment 4, the *IFT*, we cannot reject the null hypothesis of a zero effect for all comparisons. For *IFT,* we have to reject all tests. Conceptually, this is not unexpected. While for the allocation of *JT*, *RIM*, and *PS,* it is very clear that the job centers' caseworkers can decide upon (non-)allocation, for the *IFT*, that is potentially taking place at a private company, it is not solely the decision of the caseworker, but also of the company, which might have completely different objectives and information. Therefore, we decided not to investigate *IFT*, while this placebo test convinced us that for the other treatments, the CIA is credible.

# 6   Improving the allocation of the programs

Section 5 provided a list of differential effects associated with specific unemployed groups. This raises two questions regarding the observed allocation mechanism by the respective caseworkers. 1.) How well did the caseworkers allocate the programs to the unemployed? 2.) Can we improve on their assignment? In the following, we show how different hypothetical program allocations ("black box") would have worked. Further, we propose simple, transparent rules because most black-box allocations are difficult to understand. The used policy trees are easy to understand and apply and already allow to reap substantial improvements. For this, we use a random subset, men and women combined in the original shares in the population of 10,000 observations.[13] This is because, in practice, ALMP allocations are not separately done for men and women, and to get one single rule. To avoid losing the substantial differential effectiveness regarding gender, as discussed before, we add a variable for gender for the following allocations.

## 6.1 Hypothetical program allocation

We start by looking at various hypothetical allocations and comparing those to the observed allocation. For this exercise, individuals with their estimated IATEs are allocated to maximize the population's average number of days in regular employment. This is conducted by allocating those with the highest potential outcomes to the respective programs, according to given rules or restrictions. The first column of Table 6, Panel A, shows the restrictions imposed for the respective simulation. First, we investigate the allocation observed in the data. 7.20%, 5.09%, and 3.95% of all unemployed are allocated to *JT*, *RIM*, and *PS*, respectively. This results in an average of 171.25 days in regular employment within 36 months after the treatment. We take this as the benchmark value. The first simulated allocation is a purely randomized allocation, for which the resulting average days in employment are equally high as for the caseworkers' allocation. The fact that the observed allocation is only about as good as a random allocation leaves us room to improve the overall efficiency (e.g., Lechner and Smith, 2007).

The next simulation is maximizing the total benefits for every unemployed without restrictions.[14] In this scenario, only 0.09% of individuals remain untreated, while 10.21%, 22.79%, and 66.91% are assigned to *JT*, *RIM*, and *PS*, respectively. This leads to the maximum achievable average outcome of 222 days in employment.

---

[13] Since this is a 10,000 observations random subset of the used data, in which we have a 20% random subset of all non-participants, it does not represent the shares of job search and training program allocated to the whole means-tested benefits population. To account for this fact, we calculate a "gain for switchers", which is not related to the population, but to those who are reallocated by the algorithms.

[14] Starting from the observed allocation, we reallocate everyone with a higher IATE in another than the observed program/non-participation. This results in 95.49% of individuals who were switched to another program/non-participation.

*Table 6: Overall effects of simulated hypothetical program allocations*

|  | Share in different programs (in %) | | | Cum. # of days in reg. empl. in 36 months after start of treatment | Gain for switchers (in %) |
|---|---|---|---|---|---|
|  | JT | RIM | PS |  |  |
| Panel A: Hypothetical program allocations | | | | | |
| Observed | 7.20 | 5.09 | 3.95 | 171.25 | - |
| Random | 7.26 | 5.06 | 3.84 | 171.26 | + 0.00 |
| Policy simulation | | | | | |
| - No constraint | 10.21 | 22.79 | 66.91 | 222.39 | + 31.27 |
| - No constraint, only significant | 10.13 | 14.96 | 51.03 | 208.58 | + 35.54 |
| - Constrained, preference to largest gain | 7.20 | 5.09 | 3.95 | 177.35 | + 12.51 |
| - Constrained, sequential optimization | 7.20 | 5.09 | 3.95 | 178.16 | + 14.22 |
| - Constrained, preference to days since last empl. | 7.20 | 5.09 | 3.95 | 172.03 | + 1.54 |
| Panel B: Decision Trees | | | | | |
| - Constrained, 2 level | 6.58 | 4.40 | 3.82 | 172.48 | + 2.40 |
| - Constrained, 3 level | 7.35 | 4.87 | 3.85 | 174.03 | + 5.42 |
| - Constrained, 4 level | 7.33 | 4.93 | 3.72 | 174.40 | + 6.12 |

Note. – Share in non-participation can be calculated as 100% - Share JT / RIM / PS. In the "Observed" allocation, the outcome is the realized mean of cumulated days in regular employment 36 months after the start of treatment. In all the other scenarios, some individuals are hypothetically switched from the actual to another treatment status. The "gain for switchers" reflects the gains in percent for those switching the treatment status; thus, it can be higher for the same absolute gain if fewer people are reallocated. The "Constrained" allocations take the shares as in the "Observed" allocation as given to simulate a budget constraint. In Panel B, this can deviate since constraints are binding only in the training set. n level refers to the number of levels in the decision tree, i.e., 4 level (or depth 4) can have at most 16 "leaves."

For reallocating only those with IATEs statistically significantly different from zero (this additional constraint is imposed to minimize the dependence on some estimates just being positive/negative because of estimation error), we find about 61% to switch the treatment status and 39% would not have benefitted significantly from any other program (non-)participation. This results in an average outcome of 209, an increase of about 37 days more in employment on average, and for the population of switchers, an average gain of 36% (last column). According to this, most unemployed individuals should be participating in one of the training and private placement service programs if the goal is to improve chances on the labor market of as many long-term unemployed as possible. Obviously, job centers can only send a portion of the unemployed to such programs, as those are usually costly. Unfortunately, we do not have any information on the costs of the programs, which would be needed for a cost-benefit analysis. To consider the budget constraint of job centers, the following hypothetical program allocations are restricted to have the same share of participants as observed in the data so that we can talk about gains at the same costs. Bear in mind that average gains cannot be very high since about 84% of the sample remain non-participants.

Three allocation scenarios with budget constraints are presented. In the first, priority is given to those individuals with the largest gains from participating in a certain program or *non-participation*. The programs and *NP* are filled by those with the largest gains until the budget constraints for the programs or *NP* are reached. Here, the average gain would be about six days compared to the observed scenario, with an average gain for the switchers of 12%. The second allocation is reshuffling the participants to achieve a higher overall efficiency, leading to the maximum achievable outcome under budget constraints with an average of 178 days in regular employment. In a more social approach, preferring those individuals to participate in one of the programs who are in unemployment for the longest time (*days since last employment*), the average days in employment with 172 and a gain of below one day on average is marginal but still higher compared to the observed allocation.[15]

In conclusion, gains from different allocation schemes could be large, which is also true if holding the training capacity constant. Since such types of "black-box" allocations cannot be implemented transparently in the job centers' counseling process, they might not be desirable from a political point of view. Further, those mechanisms might be subject to ethical or societal concerns (Whittlestone et al., 2019; Reddy et al., 2020). This might lead to caseworkers not trusting those allocations and not following such untransparent rules. In the following subsection, we propose simple, transparent rules which are easily understandable and applicable. Further, since those rules only use a few characteristics, it is easier to decide if allocating according to them is socially and ethically acceptable.

---

[15] In this setup, we only consider individuals who have been in regular employment at least once in the past. In Table B.1, Appendix B additional results are presented showing more types of allocation rules.

## 6.2 Policy Tree

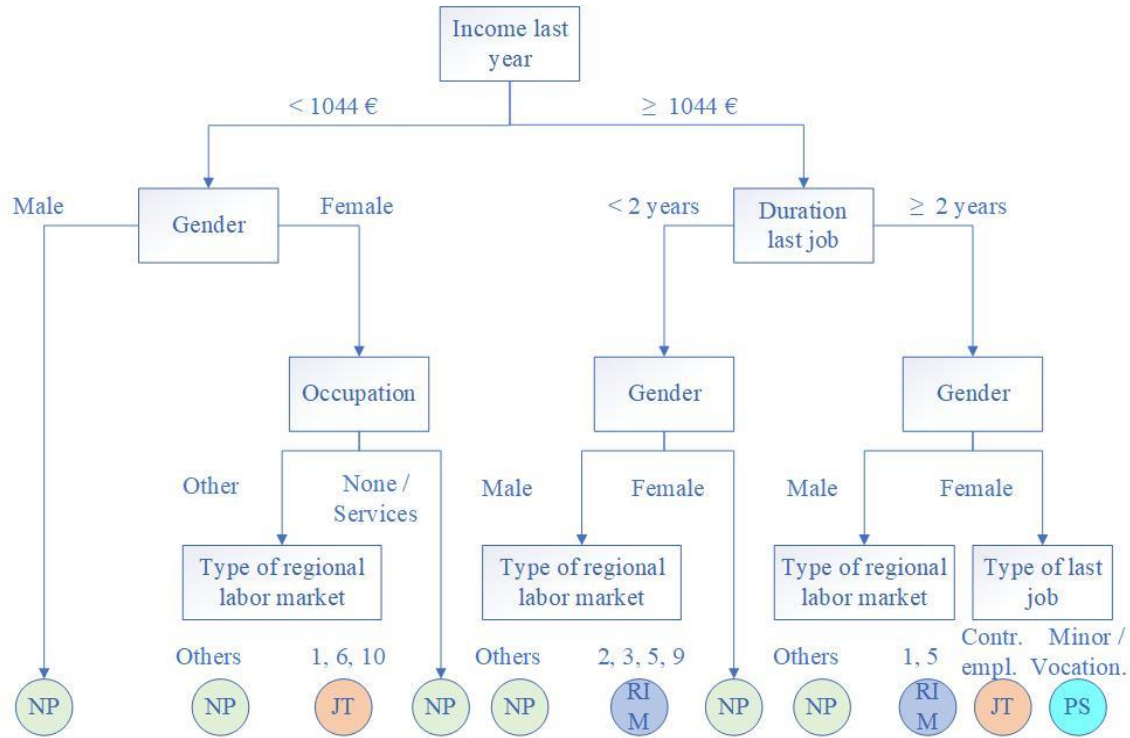

*Figure 7: Assignment rule of shallow decision tree (depth 4)*

Note. – Types of regional labor markets are 1: Cities west, average labor market situation (LMS), high GDP, high rate of long-term unemployed; 2: Cities west, above average LMS, high GDP; 3: Cities west, below average LMS, very high rate of long-term unemployed; 4: Cities, mainly east, bad LMS, very high rate of long-term unemployed; 5: Predominantly urban, west, average LMS, high rate of long-term unemployed; 6: rural, west, average LMS; 7: Predominantly urban, west and east, below average LMS; 8: rural, west, good LMS, high seasonal dynamic; 9: rural, west, very good LMS, seasonal dynamic, very low rate of long-term unemployed; 10: rural, west, very good LMS, low average rate of long-term unemployed; 11: predominantly rural, east, bad LMS, low GDP; 12: predominantly rural, east, very bad LMS, very low GDP, a high rate of long-term unemployed. NP stands for non-participation, JT for job-training, RIM for reducing impediments, and PS for placement services.

To provide easy rules for the allocation of the unemployed to the job search and training programs, we follow the approach of Zhou et al. (2022) and use shallow decision trees of depth $m = 2$ (4 strata), 3 (8 strata), and 4 (16 strata) with slight modifications of their algorithm.[16] The general idea of this policy tree is to find personalized decision rules for assignment to treatments using the characteristics of individuals that are easily comprehensible. The tree searches for this optimal decision rule in $m$ subsequent sample splits by optimally splitting in the characteristics that lead to the highest possible average outcomes for the unemployed individuals so assigned. For this exercise, we use the individual's already estimated potential outcomes for each treatment alternative.

---

[16] For details on the implementation, the interested reader refers to Online Appendix O-B. Computational restrictions do not allow for depth 5 or larger. Also, at least depth 2 (4 strata) is needed to allocate 3 treatments + non-participation.

Panel B of Table 6 summarizes how well those easy implementation rules work. Having a straightforward rule with depth 2 gains are limited to about one day. Increasing the complexity of decision trees to depths 3 and 4 results in average outcomes of about 174, implying gains of about three days on average. Figure 7 depicts our depth 4 allocation tree. At every split point, a caseworker compares the characteristics of the unemployed to the splits to be made. First, one verifies if the unemployed earned more or less than € 1,044 in the last year from income relevant to social security. If the individual is below and male, he would not be allocated to a program, while for a woman, the occupation and the type of the regional labor market play a role, etc. A smaller tree with depth 3 can be found in Appendix C, Figure C.1. Here, gender, days in regular employment in the last five years, type, length of last job, and educational achievements are used for allocation.

In general, we would like to point out four things about the allocation in Figure 7. First, as already detailed in the previous sections, gender plays a prominent role in this allocation. Men would not be allocated to the *PS*, which is the most beneficial program for them, but the gains seem to be stronger for women. Second, with past income, duration and type of last job, gender, and the type of the regional labor market, only five characteristics are used, making this a very transparent allocation scheme. Third, the presented decision tree rule strongly favors individuals living in West Germany and in regions with good labor market conditions. This might optimize the gains from the programs but could also lead to a more diverging east-west gap. Lastly, the exact budget constraint restricts the algorithm, as it must not only optimize the allocation but also restrict the sample to a certain maximum of observations allowed in the leaves relevant for the treatment allocation.[17]

# 7  Discussion

Our study analyses the effects of three different active labor market programs for the long-term unemployed, namely *job-training* (*JT*), *reducing impediments* (*RIM*), and *placement services* (*PS*). These programs are rather short in their duration but can be flexibly designed to address the needs of different types of participants. In this study, we focus on unemployed welfare recipients in Germany, a group of persons particularly in need of active support in their labor market integration. We analyzed participation effects on cumulated days in regular employment three years after the program started and propose alternative, machine learning based, allocation mechanisms.

---
[17] Results for the less restricted scenarios can be found in Appendix B, Table B.1.

The *JT*, *RIM*, and *PS* estimation results show almost no lock-in effects. These programs are generally effective for men and women in raising their cumulated days in regular employment in the first three years after treatment start. The effects are long-lasting and almost all participants benefit from the programs. This is consistent with the purpose of the programs: they successfully meet the individual needs of the unemployed and can be quite helpful to foster the integration of long-term unemployed people into the labor market. A related expectation is that the programs are more effective than their less flexible predecessors. Comparing our results with those of studies analyzing programs that were replaced by *JT*, *RIM*, and *PS* (e.g. Bernhard and Wolff, 2008; Kopf, 2013), we find that the programs we analyzed tend to lead to higher effects. Besides the general effectiveness of the investigated programs, our findings further show that individuals benefiting most from participation are characterized by an adverse labor market history. This is in line with literature attesting ALMP higher effectiveness in integrating the hard-to-place (Card et al., 2018). For women, the effects are only higher than for men when we regard *PS*; they are similarly high for *RIM* and lower for *JT*.

We find effect heterogeneity for women, but not for men. In particular, married women benefit particularly from *JT* and *PS* and hence belong to the groups with the highest potential to raise their employment rates by participation. The same holds for women living in districts with low unemployment rates and those living in West Germany, a region that was characterized by more conservative attitudes towards childcare (Boelmann et al., 2020) and scarcer provision of external childcare (Rosenfeld et al., 2004). Women in these groups again might show a relatively low labor market attachment, which implies a considerable scope of improving their employment outcomes by participation in the types of ALMP that we analyzed.

The relatively higher employment effects of *PS* among women could also originate from the labor demand side. Although the contributory employment stock increased overall by 6.1% in Germany between 2010 and 2013, some sectors like healthcare and hospitality had an above-average increase.[18] In our sample, women more often have been working in such occupations.

---

[18] Source: DataWareHouse of the Statistics Department of the German Federal Employment Agency.

Hence, private placement services might be much more effective in bringing them back to work than men who more frequently worked in other occupations with a slower rise in labor demand.

While we cannot provide a cost-benefit analysis due to missing information on costs, we find that if costs did not play a role, the means-tested benefit recipient population would benefit from more job search and training programs, as most individuals would realize positive effects. Since costs are a restriction in practice, we implemented a strict budget constraint holding the number of program participants constant and allocating individuals in certain optimized ways. Our analysis of simulated program allocations based on the IATEs showed that allocating individuals, without constraints, to maximize overall gains could raise the average program effect by more than 30%. In a feasible allocation, i.e., keeping the number of participants in each program fixed, the black-box algorithm allocates programs to increase the employment effect by 14% for those reallocated. Therefore, the programs' effectiveness could have been increased if caseworkers focused more on the groups with the highest participation effects. Furthermore, our proposed transparent allocations emphasize the importance of considering gender differences in unemployment reasons, with women having more interruptions in their employment careers due to care responsibilities and men confronting lacking job opportunities in the labor market segments they predominantly work in.

On a methodological note, we found the MCF suitable for our comprehensive analysis of the German job search and training programs. The data-driven allocation mechanisms demonstrate the power of applying machine learning methods to a policy-relevant domain. Despite investigating and analyzing the proposed allocation rules regarding which characteristics matter for a beneficial allocation of job search and training programs for means-tested benefit recipients, we show the feasibility and potential of such transparent and black-box rules. This might set the stage for politics and society to discuss whether and in what form such data-driven mechanisms should be used in practice and whether they are socially acceptable.

# Appendices

## Appendix A: Describing the individualized average treatment effects

For describing those populations of individuals benefiting most and least from ALMP participation, we show the dependence of the effects on characteristics by k-means++ clustering (Arthur and Vassilvitskii, 2007). Five clusters are formed by jointly using the IATEs of participation in one of the programs relative to non-participation. Those clusters are built by jointly sorting IATEs in increasing order to find clusters representing the group of individuals benefiting most or least. The fifth cluster represents the most affected, i.e., those with the highest treatment effects. In the first cluster, the least benefiting observations are grouped.

The means of the treatment effects by the clusters are reported in the first lines of Table A.1. This is especially interesting to see if those populations of most and least benefiting individuals differ in specific characteristics. While women from the eastern part of Germany are more present in the least benefiting population, there is no such difference for men. The picture for men and women is more uniform for classical job history characteristics. Those with worse previous labor market success, like the days since the last regular employment or the cumulated days in regular employment in the previous five years, benefit more than those with a better record. Differential effects are found for job center related characteristics. For women, a lower client-staff ratio and higher shares of sanctions are observed in the most benefiting population, whilst for men, this is observed in the least benefiting population.

*Table A.1: Descriptive statistics of clusters based on k-means clustering, IATEs*

|  | Women | | Men | |
|---|---|---|---|---|
| Cluster | Least beneficial | Most beneficial | Least beneficial | Most beneficial |
| Share of observations (in %) | 13 | 27 | 13 | 12 |
| JT vs. NP | 3 | 37 | 27 | 68 |
| RIM vs. NP | 33 | 46 | 14 | 39 |
| PS vs. NP | 8 | 49 | 7 | 55 |
| A. Personal characteristics | | | | |
| Foreigner | 0.20 | 0.25 | 0.26 | 0.22 |
| Days in regular employment (prev. 5 yrs) | 265 | 84 | 704 | 201 |
| Days since last employment | 1,798 | 2,244 | 453 | 1,758 |

*Table A.1: continued*

| | | | | |
|---|---|---|---|---|
| No vocational / academic degree | 0.49 | 0.65 | 0.47 | 0.47 |
| Vocational degree | 0.47 | 0.31 | 0.49 | 0.42 |
| Academic degree | 0.03 | 0.02 | 0.03 | 0.09 |
| Education - no schooling diploma | 0.15 | 0.23 | 0.12 | 0.11 |
| Education - secondary school | 0.40 | 0.45 | 0.50 | 0.39 |
| Education - general certificate of secondary education | 0.33 | 0.21 | 0.27 | 0.23 |
| Education - advanced technical college entrance qualification | 0.03 | 0.03 | 0.04 | 0.08 |
| Education - high school | 0.07 | 0.05 | 0.07 | 0.15 |
| Marital status - unmarried | 0.27 | 0.22 | 0.39 | 0.50 |
| Marital status - married | 0.31 | 0.39 | 0.37 | 0.28 |
| B. Job center characteristics | | | | |
| Client-staff ratio in job centers | 164 | 159 | 161 | 163 |
| Sanction intensity in job centers due to violations of duties (in %) | 0.45 | 0.62 | 0.58 | 0.53 |
| Sanction intensity in job centers due to failure in reporting (in %) | 0.63 | 0.79 | 0.75 | 0.72 |
| C. District-level characteristics | | | | |
| District unemployment rate | 11.3 | 9.6 | 9.8 | 10.5 |
| District UE rate of welfare recipients | 8.2 | 6.7 | 6.8 | 7.6 |
| Region (west=0, east=1) | 0.36 | 0.16 | 0.26 | 0.26 |

Note. – k-means++ clustering is used (Arthur and Vassilvitskii, 2007) with five clusters. Reported are clusters 1 (least beneficial) and 5 (most beneficial), while clusters 2-4 are not reported. Full results can be found in the Omline Appendix O-E. The outcome is the cumulative days in regular employment 36 months after the start of treatment.

This discrepancy is also evident with regard to local labor market conditions. While most benefiting women are rather in areas with better labor market conditions, i.e., a lower share of unemployed in general and of welfare recipients in a district, this is reversed for men. Observing personal characteristics, we find a higher percentage of foreigners, those without an academic or vocational degree, and lower educational achievements in the group of most benefiting for women but in the least benefiting group for men. Further, unmarried women are rather in the least benefiting group, while unmarried men are more present in the most benefiting group.

# Appendix B: Hypothetical program allocations, full table

*Table B.1: Overall effects of simulated hypothetical program allocations, full table*

|  | Share in different programs (in %) | | | Cum. # of days in employment | Gain for switchers (in %) |
| --- | --- | --- | --- | --- | --- |
|  | JT | RIM | PS |  |  |
| Observed | 7.20 | 5.09 | 3.95 | 171.25 | - |
| Random | 7.26 | 5.06 | 3.84 | 171.26 | + 0.00 |
| *Direct Policy Simulation* | | | | | |
| - Unconstrained | 10.21 | 22.79 | 66.91 | 222.39 | + 31.27 |
| - Unconstrained only significant | 10.13 | 14.96 | 51.03 | 208.58 | + 35.54 |
| - Constrained, highest variance | 7.20 | 5.09 | 3.95 | 175.93 | + 9.60 |
| - Constrained, preference to largest gain | 7.20 | 5.09 | 3.95 | 177.35 | + 12.51 |
| - Constrained, lowest non-participation potential outcome | 7.20 | 5.09 | 3.95 | 171.37 | + 0.23 |
| - Constrained, sequential optimization | 7.20 | 5.09 | 3.95 | 178.16 | + 14.22 |
| - Constrained, preference to days since last employment | 7.20 | 5.09 | 3.95 | 172.03 | + 1.54 |
| - Constrained, preference to highest effect relative to non-participation | 7.20 | 5.09 | 3.95 | 175.79 | + 9.27 |
| *Policy Trees* | | | | | |
| *Level 2* | | | | | |
| - Unconstrained | 0.00 | 16.91 | 83.09 | 217.46 | + 28.26 |
| - Constrained (= total number treated) | 0.00 | 0.00 | 16.25 | 176.13 | + 9.51 |
| - Constrained | 6.58 | 4.40 | 3.82 | 172.25 | + 2.40 |
| *Level 3* | | | | | |
| - Unconstrained | 0.28 | 14.98 | 84.74 | 217.64 | + 28.37 |
| - Constrained (= total number treated) | 0.00 | 0.00 | 16.17 | 176.73 | + 10.67 |
| - Constrained | 7.35 | 4.87 | 3.85 | 174.03 | + 5.42 |
| *Level 4* | | | | | |
| - Unconstrained | 0.22 | 16.34 | 83.44 | 217.86 | + 28.50 |
| - Constrained (= total number treated) | 0.54 | 0.00 | 15.72 | 177.07 | + 11.33 |
| - Constrained | 7.33 | 4.93 | 3.72 | 174.40 | + 6.12 |

Note. – Share in non-participation can be calculated as 100% - Share JT / RIM / PS. In the "Observed" allocation, the outcome is the realized mean of cumulated days in regular employment 36 months after the start of treatment. In all the other scenarios, some individuals are hypothetically switched from the actual to another treatment status. The "gain for switchers" reflects the gains in percent for those switching the

treatment status; thus, it can be higher for the same absolute gain if fewer people are reallocated. The "Constrained" allocations take the shares as in the "Observed" allocation as given to simulate a budget constraint. In Panel B, this can deviate since constraints are binding only in the training set. n level refers to the number of levels in the decision tree, i.e., 4 level (or depth 4) can have at most 16 "leaves."

# Appendix C: Optimal policy tree, depth 3

*Figure C.1: Assignment rule of shallow decision tree (depth 3)*

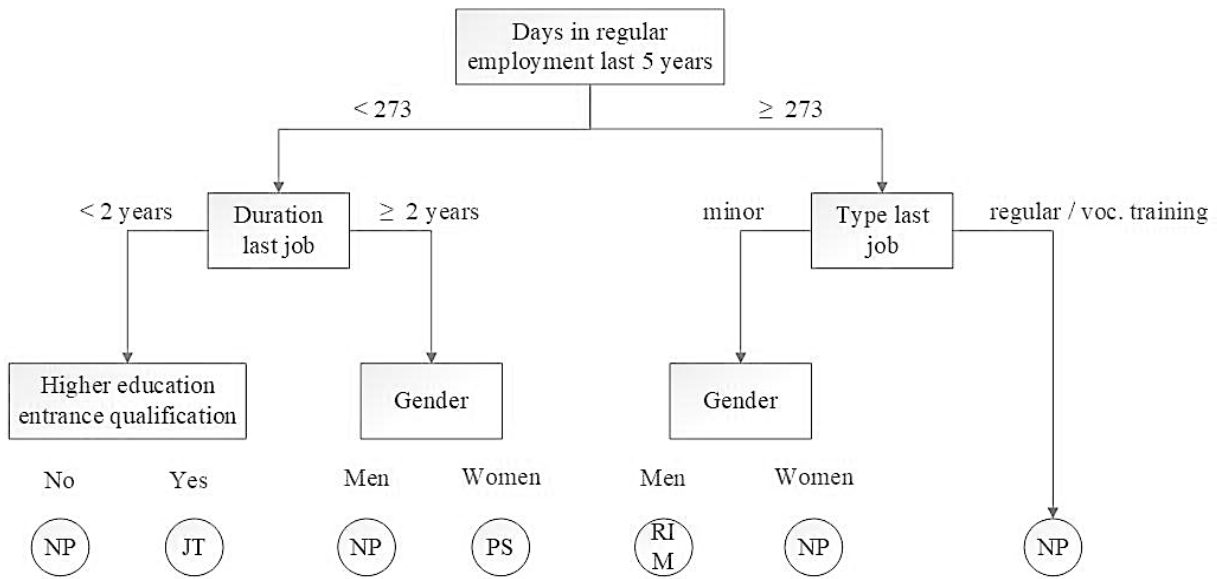

Note. – NP: non-participation, JT: job-training, RIM: reducing impediments, PS: placement services.

# Online Appendices

## Online Appendix O-A: Additional Results on the group average level

*Figure O-A.1: Difference of GATEs to ATE by the district unemployment rate*

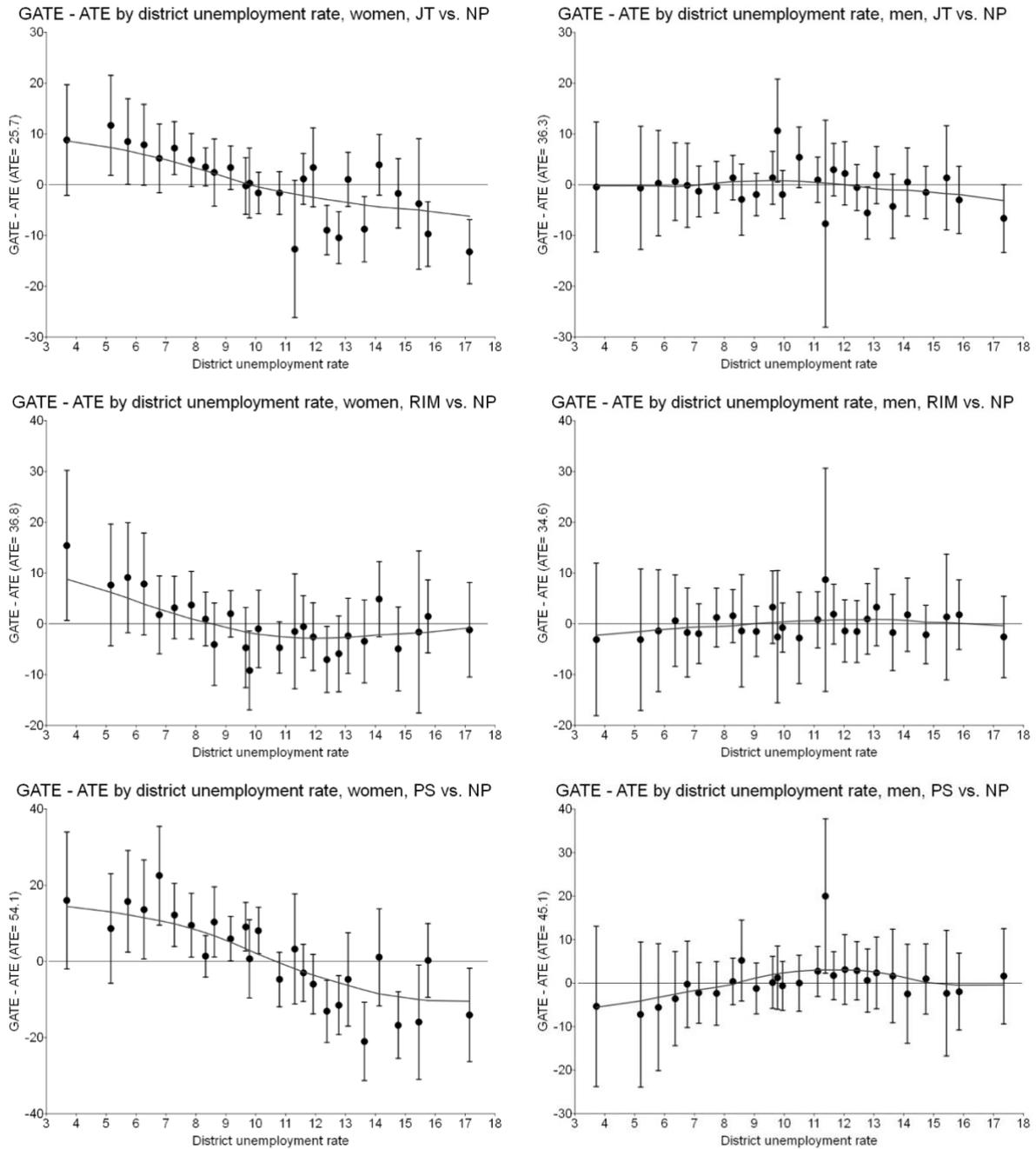

Note. – 90% confidence intervals. ATE estimates are in parentheses on the left-hand side of the graphs. The outcome is cumulative days in regular employment 36 months after the start of treatment. The solid (green) line represents the (kernel-)smoothed GATE estimates and is for illustration purposes only. NP: non-participation, JT: job-training, RIM: reducing impediments, PS: placement services.

*Figure O-A.2: Difference of GATEs to ATE by the district unemployment rate of welfare recipients*

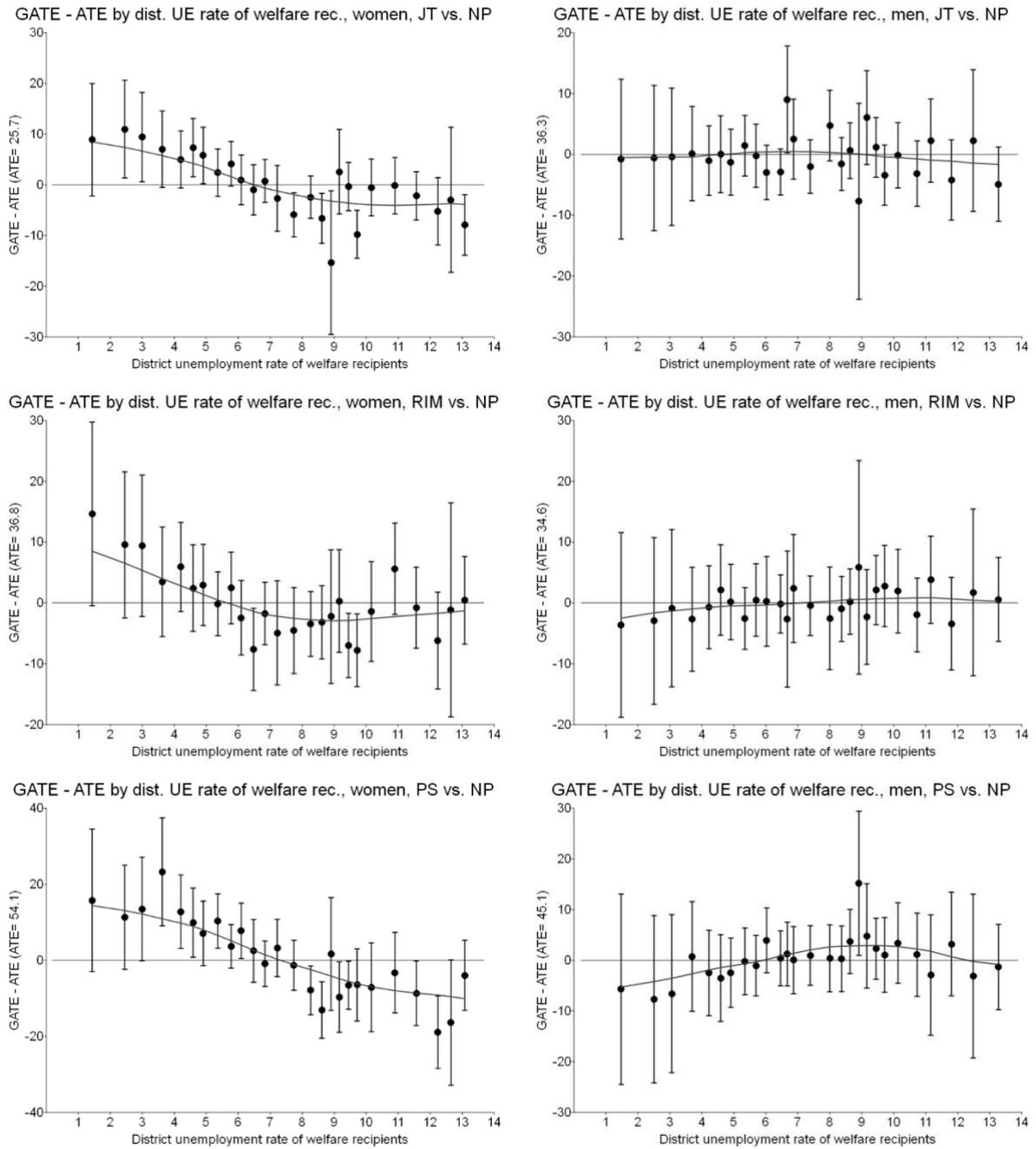

Note. – 90% confidence intervals. ATE estimates are in parentheses on the left-hand side of the graphs. The outcome is cumulative days in regular employment 36 months after the start of treatment. The solid (green) line represents the (kernel-)smoothed GATE estimates and is for illustration purposes only. NP: non-participation, JT: job-training, RIM: reducing impediments, PS: placement services.

*Figure O-A.3: Difference of GATEs to ATE by sanction intensity (violation of duties) in job center*

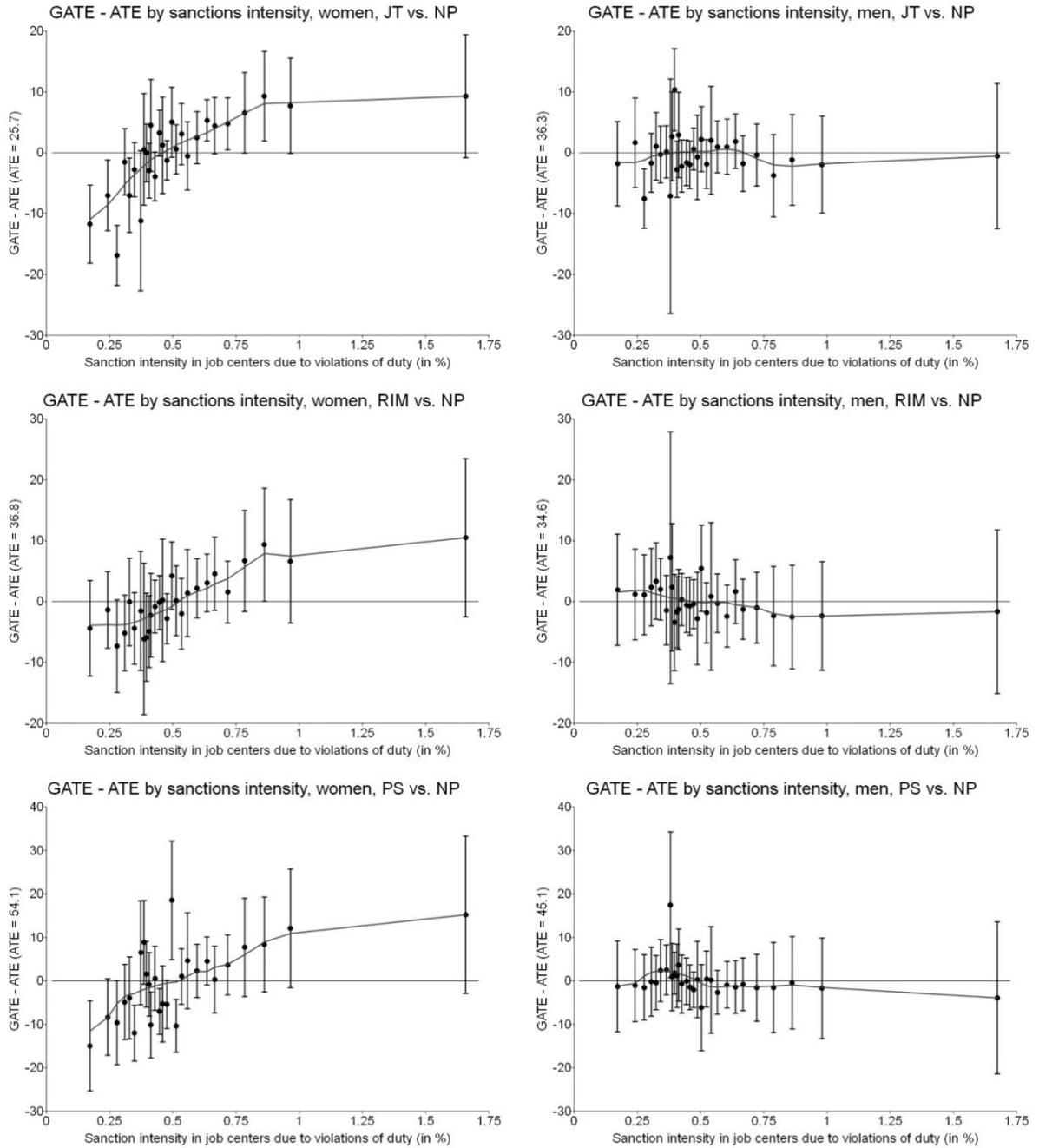

Note. – 90% confidence intervals. ATE estimates are in parentheses on the left-hand side of the graphs. The outcome is cumulative days in regular employment 36 months after the start of treatment. The solid (green) line represents the (kernel-)smoothed GATE estimates and is for illustration purposes only. NP: non-participation, JT: job-training, RIM: reducing impediments, PS: placement services.

*Figure O-A.4: Difference of GATEs to ATE by cumulative days in regular employment in the last five years*

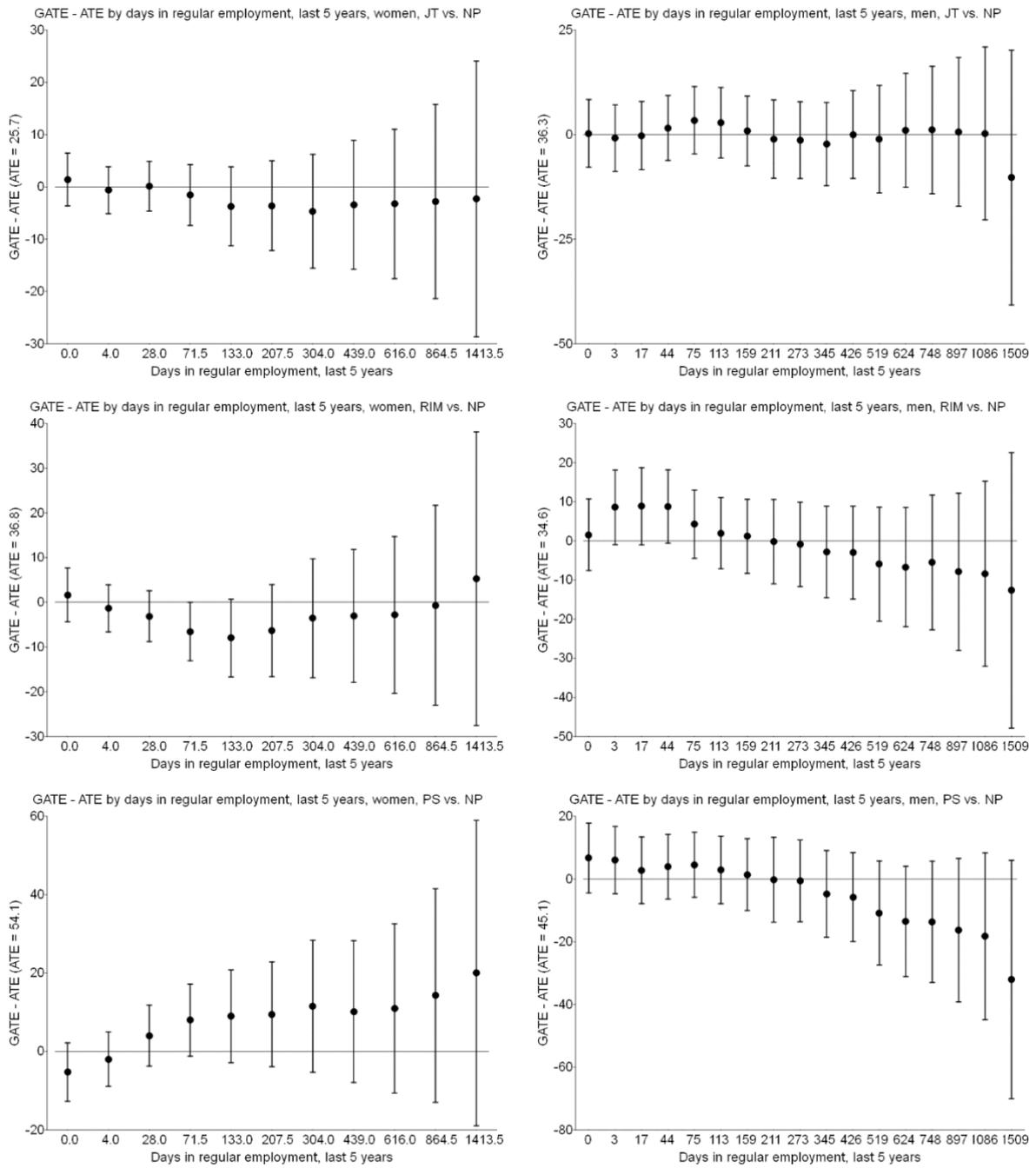

Note. – 90% confidence intervals. ATE estimates are in parentheses on the left-hand side of the graphs. The outcome is cumulative days in regular employment 36 months after the start of treatment. The solid (green) line represents the (kernel-)smoothed GATE estimates and is for illustration purposes only. NP: non-participation, JT: job-training, RIM: reducing impediments, PS: placement services.

*Figure O-A.5: Difference of GATEs to ATE by age*

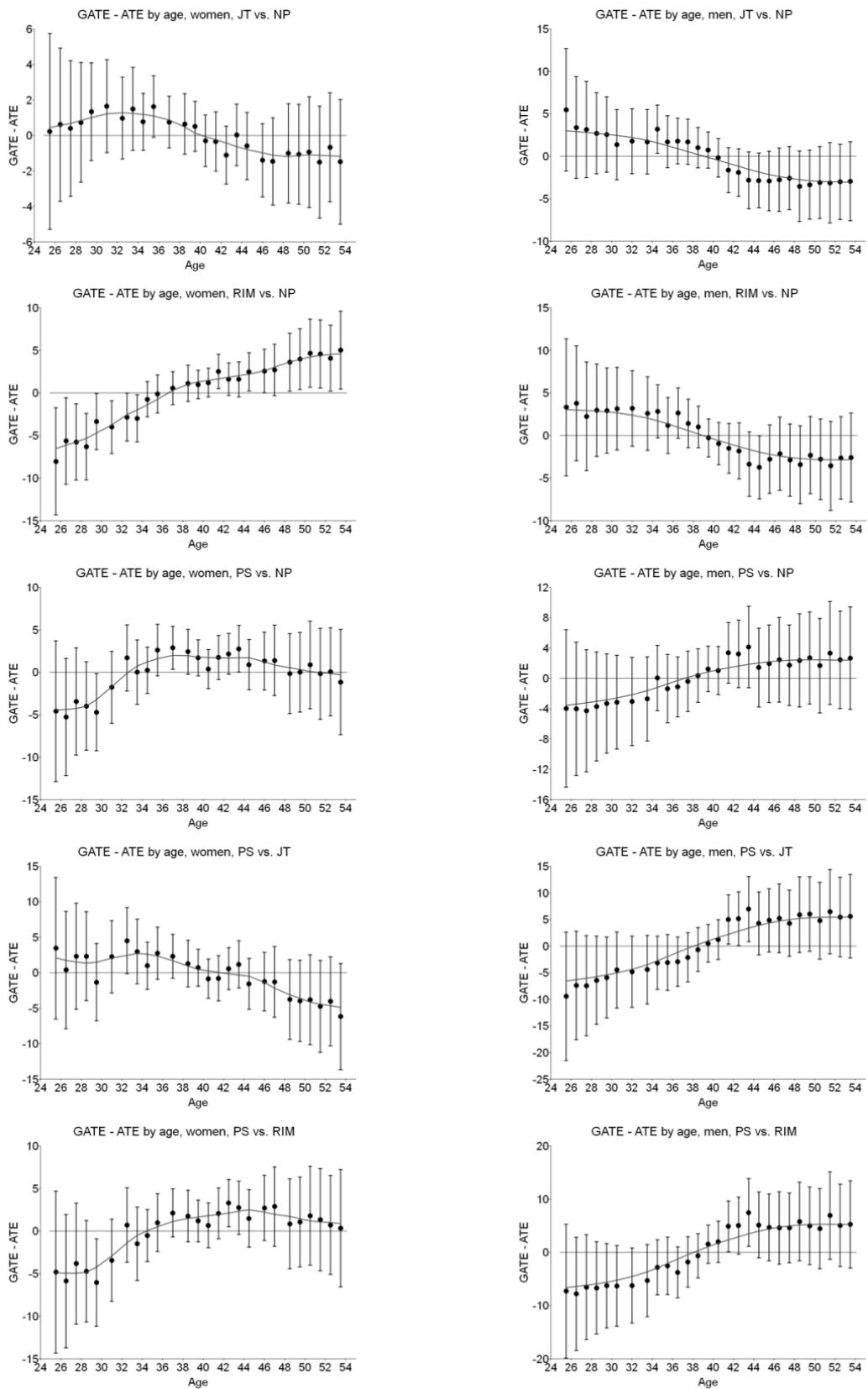

Note. – 90% confidence intervals. ATE estimates are in parentheses on the left-hand side of the graphs. The outcome is cumulative days in regular employment 36 months after the start of treatment. The solid (green) line represents the (kernel-)smoothed GATE estimates and is for illustration purposes only. NP: non-participation, JT: job-training, RIM: reducing impediments, PS: placement services.

*Figure O-A.6: Difference of GATEs to ATE by family status*

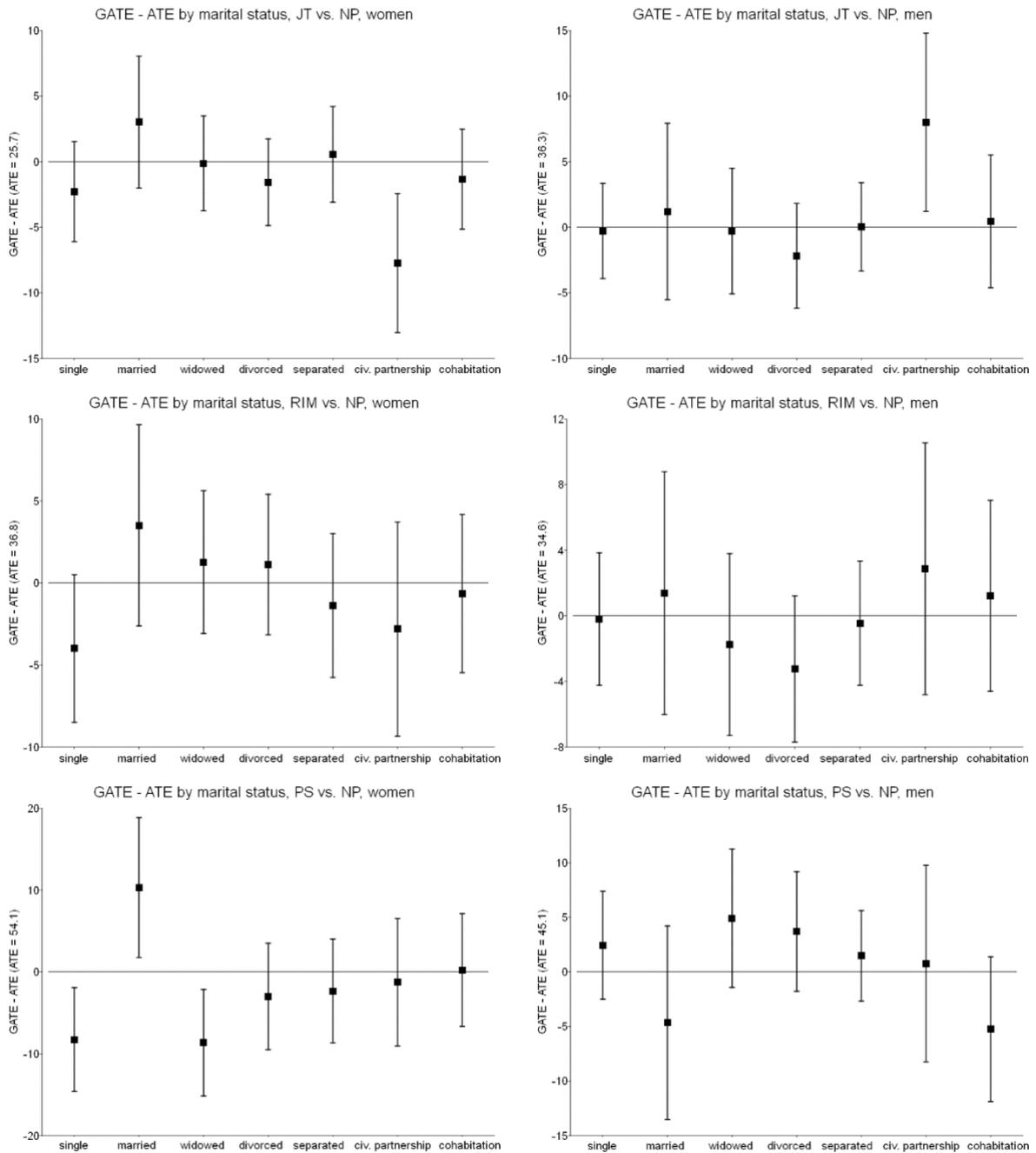

Note. – 90% confidence intervals. ATE estimates are in parentheses on the left-hand side of the graphs. The outcome is cumulative days in regular employment 36 months after the start of treatment. NP: non-participation, JT: job-training, RIM: reducing impediments, PS: placement services.

*Figure O-A.7: Difference of GATEs to ATE by job center client-staff ratio*

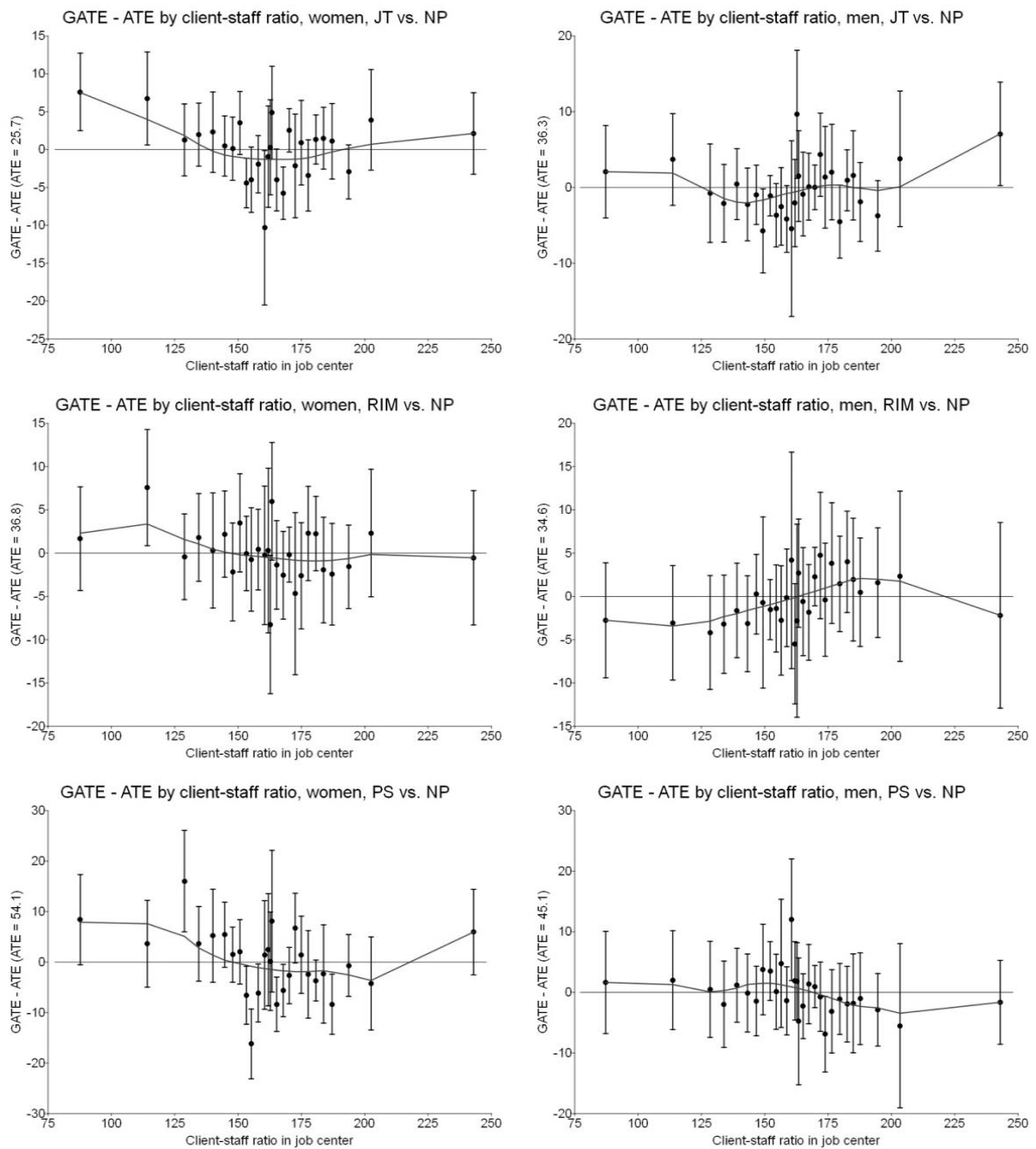

Note. – 90% confidence intervals. ATE estimates are in parentheses on the left-hand side of the graphs. The outcome is cumulative days in regular employment 36 months after the start of treatment. The solid (green) line represents the (kernel-)smoothed GATE estimates and is for illustration purposes only. NP: non-participation, JT: job-training, RIM: reducing impediments, PS: placement services.

*Figure O-A.8: Difference of GATEs to ATE by educational achievement*

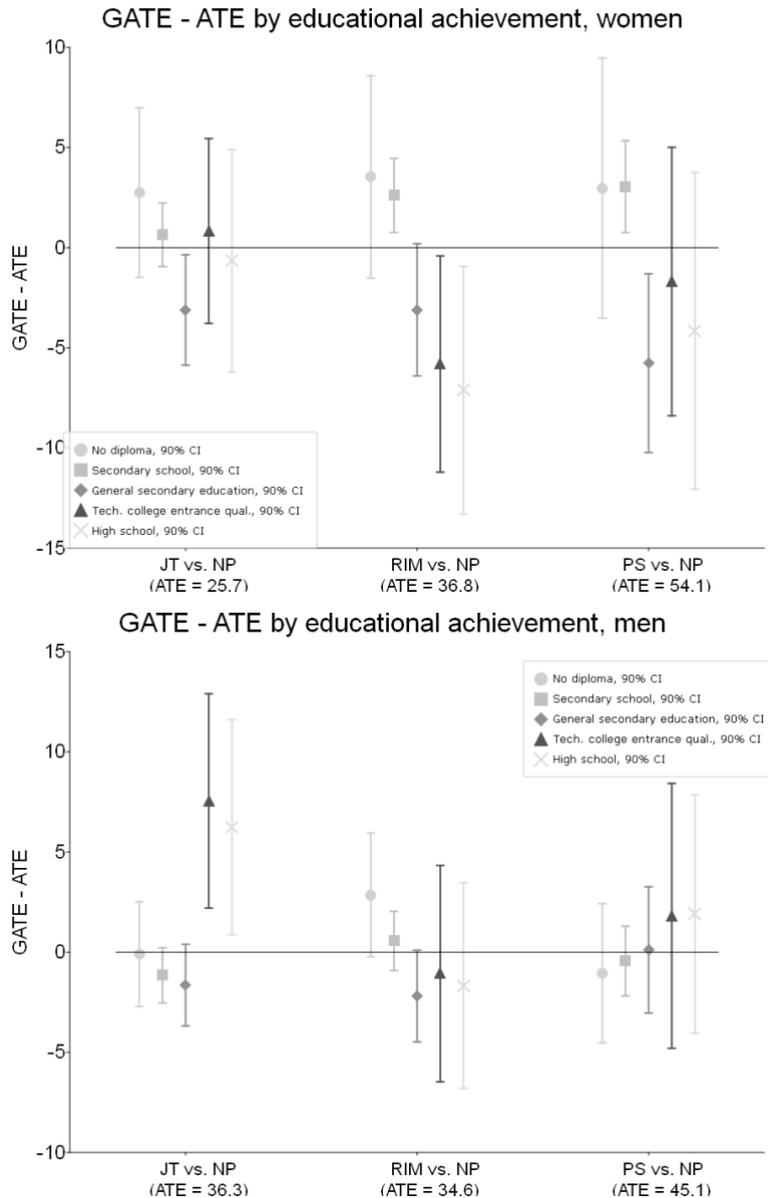

Note. – 90% confidence intervals (CI). ATE estimates are in parentheses at the bottom of the graphs. The outcome is the cumulative days in regular employment in 36 months after the start of treatment. NP: non-participation, JT: job-training, RIM: reducing impediments, PS: placement services.

# Online Appendix O-B: Implementation of Shallow Decision Trees

This study uses the same implementation of shallow decision trees as Cockx et al. (2023), and their Appendix B.4, provide the concrete algorithms described in the following. The general idea goes back to Zhou et al. (2022), and their algorithm 2, is modified in three aspects to fit our purpose.

First, we allow for unordered categorical variables. Second, on higher levels of the tree, we use a finer grid for computing possible splitting rules.[19] Third, to take budget constraints regarding training program shares into account, constraints are enforced to ensure overall program shares and maximum individual program shares. Being able to work with unordered categorical variables is especially useful in cases when the number of potential splits is limited, like the underlying. A single split on a categorical variable is more informative than any resulting binary variable used if it is impossible to work with categorical variables. Having implemented this ensures that we do not favor any variable over another.

By using a finer grid at higher levels of the tree, we expect to have higher precision at those splits. Since more data are available at lower levels of the tree, the grid can be a bit coarser to save some computation time. This adaptive way should increase precision over the one-size-fits-all approach previously proposed.

The budget constraints are implemented out of necessity due to the nature of the problem at hand. First, it is useless for policy analysis to provide allocations with arbitrary numbers of participants in the programs. Second, while it would be possible to work with a monetary budget constraint, we do not know the costs of each program, just as we do not know whether the capacities for the respective programs could be expanded at will. For these reasons, we implement the rule that, at most, as many as are observed in the sample may be in the programs after the reallocation. The interested reader is referred to Appendix B.4 in Cockx et al. (2023) for more technical and implementation details.

## Online Appendix O-C: MCF implementation and feature selection

### Online Appendix O-C.1: Practical implementation

Our overall sample consists of 302,626 observations, as discussed in Section 3. We do the estimations using samples of women and men separately for two reasons. First, we expect different effects with regard to gender, and second, for computational reasons.

The respective subpopulations are randomly divided into proportions of 75% for training and 25% for predicting the various causal effects. From the training data, 20% of the observations are taken for a feature selection procedure to reduce the extensive set of potential confounders to a smaller, most relevant set of covariates. The detailed motivation and introduction to this procedure is discussed in Online Appendix O-C.2 on feature selection.

---

[19] $4^{th}$ level, i.e., at the top of the tree, A/8; $3^{rd}$ level A/4; $2^{nd}$ level A/2; and the $1^{st}$ level, at the bottom of the tree, A; with A being the approximation parameter as also used in Zhou et al. (2022) as one single global approximation level.

Tuning parameters, like the minimum leaf size and the number of features available in each split, are determined in a grid search by out-of-bag minimizing the MSE. The share of subsampling is fixed to 2/3 of training observations, and the number of trees is set to 1,000. Several choices are varied to investigate the implied sensitivity, while the conclusions drawn are unaffected.

Online Appendix O-C.2: Feature selection

In causal studies, having too few (relevant) covariates might lead to omitted variable bias. Having too many (irrelevant) covariates might lead to not controlling for the essential, confounding variables in tree-based methods.

One needs to be very careful when implementing some variable selection procedure as a pre-step in causal studies not to omit variables that are related even mildly to both the treatment selection and the outcome. Still, e.g., Borup et al. (2023) find that excluding weak predictors before a random forest estimation improves the tree's strength. In the application of forest-type estimators, this predictor targeting is in use, using LASSO (Kotchoni et al., 2019) or Elastic Net (Borup and Schütte, 2022; Bork et al., 2020) to pre-select features. Since it is unclear in our setup which functional forms of covariates to include in a LASSO or Elastic Net and how to use this to detect all relevant confounders influencing both the potential outcome and the treatment allocation, we chose to conduct another approach.

First, we estimate a MCF on the 20% subsample and check the influence of all the variables on the estimates in the form of a variable importance measure. For the final analysis on the separate 80% subsample, we only use those covariates with positive variable importance. This is a very conservative choice, which results in a final set of about 70% of all covariates, i.e., about 140. Since there is no test to check if all the relevant confounders are included in the final set of covariates, we investigated two things. First, we checked if all the essential confounders, as described and discussed in the identification chapter (Section 4.2), survived the selection procedure. Second, it is crucial to check if the estimates change substantially compared to an estimation without using a pre-feature selection procedure. In case of discrepancies in the results, this might point to omitted variables. We checked this for the main results, and since the results are consistent, we chose to use this computationally more attractive procedure, including the pre-feature selection.

# Online Appendix O-D: Full descriptive statistics

*Table O-D.1: Summary of covariates, men*

| | Men | | | |
|---|---|---|---|---|
| Variable | Job-training | Reducing impediments | Placement services | Non-participants |
| Panel A: Sociodemographic characteristics | | | | |
| Age at sampling date (in years) | 38.28 | 37.74 | 39.03 | 39.70 |
| | (8.66) | (8.36) | (39.03) | (8.57) |
| Region (west=0, east=1) | 0.29 | 0.26 | 0.14 | 0.35 |
| Foreigner | 0.20 | 0.21 | 0.29 | 0.20 |
| Nationality - Germany | 0.80 | 0.79 | 0.71 | 0.80 |
| Nationality - EU (w/o Germany, w/o former YUG) | 0.03 | 0.04 | 0.05 | 0.03 |
| Nationality - Europe rest (w/o former YUG) | 0.02 | 0.02 | 0.03 | 0.02 |
| Nationality - Turkey | 0.07 | 0.07 | 0.11 | 0.07 |
| Nationality - former Soviet Union (without EU members) | 0.02 | 0.02 | 0.02 | 0.02 |
| Nationality - other countries | 0.05 | 0.06 | 0.08 | 0.05 |
| Continent - Germany | 0.80 | 0.79 | 0.71 | 0.80 |
| Continent - Europe | 0.15 | 0.15 | 0.21 | 0.14 |
| Continent - Africa | 0.02 | 0.02 | 0.03 | 0.02 |
| Continent - America | 0.00 | 0.00 | 0.00 | 0.00 |
| Continent - Asia / Oceania | 0.03 | 0.03 | 0.04 | 0.04 |
| Continent - stateless / unknown | 0.00 | 0.00 | 0.00 | 0.00 |
| Education - no schooling diploma | 0.12 | 0.13 | 0.14 | 0.14 |
| Education - secondary school | 0.49 | 0.51 | 0.48 | 0.48 |
| Education - general certificate of secondary education | 0.25 | 0.25 | 0.21 | 0.26 |
| Education - advanced technical college entrance qualification | 0.04 | 0.04 | 0.05 | 0.04 |
| Education - high school | 0.08 | 0.07 | 0.12 | 0.08 |
| Education - missing values | 0.01 | 0.01 | 0.01 | 0.01 |
| No vocational / academic degree | 0.49 | 0.51 | 0.54 | 0.50 |
| Vocational degree | 0.47 | 0.46 | 0.41 | 0.46 |
| Academic degree | 0.04 | 0.03 | 0.05 | 0.04 |
| Vocational degree - missing values | 0.00 | 0.00 | 0.00 | 0.00 |
| Family status - single | 0.51 | 0.51 | 0.45 | 0.48 |
| Family status - married | 0.26 | 0.26 | 0.32 | 0.27 |
| Family status - no longer married | 0.15 | 0.15 | 0.17 | 0.17 |
| Family status - cohabitation | 0.08 | 0.09 | 0.07 | 0.08 |
| Marital status - unmarried | 0.51 | 0.51 | 0.45 | 0.48 |
| Marital status - married | 0.26 | 0.26 | 0.32 | 0.27 |
| Marital status - widowed | 0.00 | 0.00 | 0.00 | 0.00 |
| Marital status - divorced | 0.10 | 0.09 | 0.10 | 0.11 |
| Marital status - separated | 0.05 | 0.05 | 0.06 | 0.06 |
| Marital status - cohabitation | 0.08 | 0.09 | 0.07 | 0.08 |
| BG type - single | 0.62 | 0.61 | 0.58 | 0.61 |
| BG type - single with adult aged <25 years | 0.00 | 0.00 | 0.00 | 0.00 |
| BG type - single with children aged <18 years | 0.02 | 0.01 | 0.01 | 0.02 |
| BG type - single with children aged <18 years and adult aged <25 years | 0.00 | 0.00 | 0.00 | 0.00 |
| BG type - couple w/o children | 0.08 | 0.08 | 0.08 | 0.09 |
| BG type - couple w/o children, but with adult aged <25 years | 0.02 | 0.02 | 0.02 | 0.02 |
| BG type - couple with children aged <18 years | 0.22 | 0.24 | 0.27 | 0.22 |
| BG type - couple with children aged <18 years and adult aged <25 years | 0.03 | 0.03 | 0.03 | 0.03 |
| BG type - other | 0.01 | 0.01 | 0.00 | 0.01 |
| Pregnancy - neither sample member nor partner | 0.98 | 0.98 | 0.98 | 0.98 |
| Pregnancy - sample member | 0.00 | 0.00 | 0.00 | 0.00 |
| Pregnancy - partner | 0.02 | 0.02 | 0.02 | 0.02 |
| Pregnancy - missing value sample member | 0.00 | 0.00 | 0.00 | 0.00 |
| Pregnancy - missing value partner | 0.00 | 0.00 | 0.00 | 0.00 |
| Partner living in HH (1=yes) | 0.34 | 0.36 | 0.39 | 0.35 |
| Child in HH (1=yes) | 0.27 | 0.28 | 0.31 | 0.27 |
| Own child in HH (1=yes) | 0.27 | 0.28 | 0.31 | 0.27 |
| Profile of unemployed - Market | 0.04 | 0.03 | 0.07 | 0.03 |
| Profile of unemployed - Activation | 0.03 | 0.02 | 0.05 | 0.02 |
| Profile of unemployed - Support | 0.23 | 0.25 | 0.29 | 0.21 |
| Profile of unemployed - Development | 0.38 | 0.41 | 0.37 | 0.38 |
| Profile of unemployed - Stabilization | 0.07 | 0.07 | 0.05 | 0.08 |
| Profile of unemployed - Assistance | 0.15 | 0.14 | 0.11 | 0.21 |
| Profile of unemployed - integrated, but dependent on help | 0.03 | 0.03 | 0.03 | 0.02 |
| Profile of unemployed - not specified | 0.03 | 0.02 | 0.02 | 0.02 |
| Profile of unemployed - missing | 0.04 | 0.02 | 0.02 | 0.01 |
| Profile of unemployed - assignment not required | 0.01 | 0.01 | 0.01 | 0.01 |
| Panel B: Labor market history | | | | |
| Cumul. days in unemployment in the last year | 254.80 (112.75) | 262.46 (105.58) | 265.59 (103.94) | 277.17 (101.81) |
| Cumul. days in unemployment in the last year - 1 month | 0.05 | 0.03 | 0.02 | 0.02 |
| Cumul. days in unemployment in the last year - >1-2 months | 0.04 | 0.03 | 0.03 | 0.02 |
| Cumul. days in unemployment in the last year - >2-3 months | 0.05 | 0.04 | 0.04 | 0.04 |
| Cumul. days in unemployment in the last year - >3-6 months | 0.14 | 0.15 | 0.15 | 0.14 |
| Cumul. days in unemployment in the last year - >6-9 months | 0.19 | 0.19 | 0.20 | 0.17 |
| Cumul. days in unemployment in the last year - >9-12 months | 0.54 | 0.56 | 0.56 | 0.62 |
| Cumul. days in unemployment in the last 2 years | 438.66 (213.68) | 451.96 (202.48) | 456.77 (202.70) | 488.99 (202.01) |
| Cumul. days in unemployment in the last 5 years | 977.01 (496.19) | 1,001.77 (470.91) | 992.66 (481.31) | 1,081.21 (479.48) |
| Cumul. days in unemployment in the last 5 years - 0-6 months | 0.06 | 0.04 | 0.04 | 0.04 |
| Cumul. days in unemployment in the last 5 years - >6-12 months | 0.07 | 0.06 | 0.07 | 0.05 |
| Cumul. days in unemployment in the last 5 years - >12-24 months | 0.20 | 0.20 | 0.21 | 0.17 |
| Cumul. days in unemployment in the last 5 years - >24-36 months | 0.23 | 0.25 | 0.24 | 0.23 |
| Cumul. days in unemployment in the last 5 years - >36-48 months | 0.24 | 0.24 | 0.24 | 0.25 |
| Cumul. days in unemployment in the last 5 years - >48 months | 0.20 | 0.20 | 0.20 | 0.26 |
| Cumul. days in job search in the last 1 year | 310.75 (97.34) | 322.24 (83.82) | 324.06 (80.33) | 328.24 (76.55) |
| Cumul. days in job search in the last 2 years | 566.90 (210.70) | 588.28 (189.95) | 588.14 (188.79) | 608.77 (178.63) |
| Cumul. days in job search in the last 5 years | 1,285.18 (531.51) | 1,329.59 (492.01) | 1,295.50 (504.48) | 1,374.14 (480.72) |
| Cumul. days in UB II in the last 1 year | 288.75 (119.52) | 303.13 (105.56) | 302.30 (104.19) | 318.32 (91.27) |
| Cumul. days in UB II in the last 1 year - 1 month | 0.07 | 0.04 | 0.03 | 0.02 |
| Cumul. days in UB II in the last 1 year - >1-2 months | 0.04 | 0.03 | 0.03 | 0.02 |
| Cumul. days in UB II in the last 1 year - >2-6 months | 0.09 | 0.09 | 0.10 | 0.08 |
| Cumul. days in UB II in the last 1 year - >6-9 months | 0.08 | 0.08 | 0.09 | 0.07 |
| Cumul. days in UB II in the last 1 year - >9-12 months | 0.72 | 0.76 | 0.75 | 0.81 |
| Cumul. days in UB II in the last 2 years | 534.42 | 558.00 | 550.75 | 593.15 |
| | (252.86) | (232.95) | (232.95) | (210.63) |
| Cumul. days in UB II in the last 5 years | 1,196.07 | 1,243.57 | 1,204.96 | 1,312.36 |
| | (623.13) | (587.76) | (595.77) | (561.15) |
| Cumul. days in UB II in the last 5 years - 0-6 months | 0.10 | 0.07 | 0.07 | 0.05 |
| Cumul. days in UB II in the last 5 years - >6-12 months | 0.06 | 0.06 | 0.07 | 0.05 |
| Cumul. days in UB II in the last 5 years - >12-24 months | 0.11 | 0.11 | 0.11 | 0.10 |
| Cumul. days in UB II in the last 5 years - >24-36 months | 0.12 | 0.12 | 0.12 | 0.11 |
| Cumul. days in UB II in the last 5 years - >36-48 months | 0.14 | 0.15 | 0.15 | 0.15 |
| Cumul. days in UB II in the last 5 years - >48 months | 0.48 | 0.50 | 0.47 | 0.55 |
| Experience in UB I (1=yes) | 0.05 | 0.05 | 0.05 | 0.04 |
| Cumul. days in UB I in the last 1 year | 34.11 | 34.24 | 36.33 | 23.82 |
| | (82.44) | (81.55) | (83.26) | (68.58) |
| Cumul. days in UB I in the last 2 years | 56.38 | 57.51 | 63.86 | 44.03 |
| | (114.19) | (114.73) | (120.22) | (102.70) |
| Cumul. days in UB I in the last 5 years | 121.36 | 122.91 | 136.87 | 106.14 |
| | (177.53) | (178.32) | (183.50) | (170.44) |
| Days since last employment | 1,527.86 | 1,488.66 | 1,383.71 | 1,821.92 |
| | (1,612.28) | (1,590.92) | (1,579.42) | (1,774.76) |
| Cumul. days in regular employment in the last year | 23.61 | 22.51 | 26.90 | 17.34 |
| | (61.02) | (58.58) | (62.00) | (51.81) |
| Cumul. days in regular employment in the last 5 years | 285.30 | 282.00 | 319.94 | 224.59 |
| | (403.41) | (394.55) | (402.39) | (357.51) |
| Cumul. days in contributory employment in the last year | 27.43 | 25.96 | 29.91 | 21.14 |
| | (65.40) | (62.58) | (64.85) | (56.85) |
| Cumul. days in contributory employment in the last 2 years | 93.99 | 94.10 | 102.34 | 72.49 |
| | (150.86) | (149.32) | (149.35) | (132.90) |
| Cumul. days in contributory employment in the last 5 years | 288.55 | 286.83 | 325.36 | 236.24 |
| | (391.30) | (382.41) | (394.45) | (350.93) |
| Cumul. days in vocational training in the last year | 1.04 | 0.77 | 0.80 | 0.61 |
| | (14.59) | (11.61) | (12.11) | (10.89) |
| Cumul. days in vocational training in the last 2 years | 4.09 | 3.73 | 3.33 | 2.53 |
| | (41.14) | (37.29) | (37.47) | (31.73) |
| Cumul. days in vocational training in the last 5 years | 24.03 | 23.62 | 18.07 | 15.41 |
| | (131.68) | (129.31) | (114.27) | (104.50) |
| Cumul. days in minor employment in the last year | 41.79 | 39.81 | 56.67 | 48.65 |
| | (98.21) | (95.02) | (112.20) | (106.00) |
| Cumul. days in minor employment in the last 2 years | 79.41 | 75.16 | 102.13 | 89.29 |
| | (174.81) | (167.62) | (194.78) | (186.59) |
| Cumul. days in minor employment in the last 5 years | 175.01 | 165.53 | 213.57 | 187.71 |
| | (342.86) | (328.61) | (375.57) | (361.53) |
| Cumul. days in bridging allowance in the last 5 years | 1.42 | 1.57 | 1.62 | 1.55 |
| | (15.35) | (16.16) | (16.47) | (16.11) |
| Cumul. days in One-Euro-Jobs in the last year | 16.80 | 18.73 | 12.89 | 15.54 |
| | (53.03) | (55.86) | (46.48) | (50.28) |
| | 36.54 | 41.30 | 29.87 | 36.09 |
| Cumul. days in One-Euro-Jobs in the last 2 years | (90.85) | (96.88) | (81.77) | (90.74) |
| Cumul. days in One-Euro-Jobs in the last 5 years | 91.42 | 101.30 | 75.91 | 91.08 |
| | (173.52) | (180.90) | (152.56) | (172.43) |
| Cumul. days in subsidized public employment in the last year | 2.35 | 2.55 | 1.84 | 2.90 |
| | (20.47) | (20.78) | (17.25) | (22.47) |
| Cumul. days in subsidized public employment in the last 2 years | 6.04 | 6.87 | 4.48 | 7.42 |
| | (36.83) | (38.42) | (31.48) | (41.23) |
| Cumul. days in subsidized public employment in the last 5 years | 14.91 | 16.69 | 11.74 | 17.16 |
| | (61.00) | (65.02) | (56.35) | (68.18) |
| Cumul. days in subsidized employment in the last year | 2.24 | 1.73 | 1.86 | 1.39 |
| | (17.23) | (15.21) | (14.84) | (13.19) |
| Cumul. days in subsidized employment in the last 5 years | 12.59 | 12.63 | 12.27 | 10.47 |
| | (46.70) | (48.99) | (45.72) | (44.16) |
| Experience in further vocational training (1=yes) | 0.35 | 0.37 | 0.32 | 0.33 |
| Cumul. days in further vocational training in the last 1 year | 8.05 | 6.73 | 9.11 | 7.16 |
| | (33.80) | (30.39) | (36.65) | (32.08) |
| Cumul. days in further vocational training in the last 5 years | 30.64 | 30.52 | 31.27 | 26.69 |
| | (94.77) | (95.67) | (96.59) | (89.53) |
| Experience in in-firm SAI (1=yes) | 0.05 | 0.05 | 0.04 | 0.04 |
| Cumul. days in in-firm SAI in the last year | 0.52 | 0.52 | 0.48 | 0.43 |
| | (3.12) | (3.02) | (3.07) | (2.86) |
| Cumul. days in in-firm SAI in the last 2 years | 0.52 | 0.52 | 0.48 | 0.43 |
| | (3.12) | (3.02) | (3.07) | (2.86) |
| Experience in classroom SAI (1=yes) | 0.11 | 0.14 | 0.11 | 0.07 |
| Cumul. days in classroom SAI in the last year | 3.85 | 5.74 | 6.38 | 3.04 |
| | (18.90) | (21.52) | (26.53) | (17.37) |
| Cumul. days in classroom SAI in the last 2 years | 3.85 | 5.74 | 6.38 | 3.04 |
| | (18.90) | (21.52) | (26.53) | (17.37) |
| Experience in *job-training* (*JT*) (1=yes) | 0.07 | 0.05 | 0.04 | 0.03 |
| Experience in *reducing impediments* (*RIM*) (1=yes) | 0.02 | 0.07 | 0.02 | 0.02 |
| Cumul. days in SAI in the last year | 4.35 | 6.21 | 6.82 | 3.46 |
| | (19.06) | (21.50) | (26.68) | (17.55) |
| Cumul. days in SAI in the last 2 years | 4.35 | 6.21 | 6.82 | 3.46 |
| | (19.06) | (21.50) | (26.68) | (17.55) |
| Experience in in-firm training (1=yes) | 0.23 | 0.23 | 0.19 | 0.20 |
| Cumul. days in in-firm training in the last year | 0.55 | 0.54 | 0.50 | 0.45 |
| | (3.26) | (3.26) | (3.21) | (3.02) |
| Cumul. days in in-firm training in the last 5 years | 3.25 | 3.46 | 2.77 | 2.89 |
| | (10.67) | (11.03) | (10.15) | (10.21) |
| Experience in classroom training (1=yes) | 0.57 | 0.64 | 0.57 | 0.49 |
| Cumul. days in classroom training in the last year | 6.29 | 9.33 | 9.59 | 4.81 |
| | (22.07) | (25.94) | (30.30) | (20.09) |
| Cumul. days in classroom training in last 5 years | 17.42 | 23.72 | 19.96 | 13.33 |
| | (34.73) | (42.19) | (44.90) | (31.41) |
| Ever received mild sanction (1=yes) | 0.16 | 0.16 | 0.14 | 0.17 |
| Cumul. days in mild sanctions in the last year | 6.89 | 6.62 | 5.96 | 7.66 |
| | (27.42) | (26.25) | (25.49) | (29.41) |
| Cumul. days in mild sanctions in the last year - never sanctioned | 0.92 | 0.92 | 0.93 | 0.92 |
| Cumul. days in mild sanctions in the last year - once (3 months) | 0.06 | 0.06 | 0.06 | 0.06 |
| Cumul. days in mild sanctions in the last year - more than once | 0.02 | 0.02 | 0.01 | 0.02 |
| Ever received strong sanction (1=yes) | 0.23 | 0.24 | 0.21 | 0.21 |
| Cumul. days in strong sanctions in the last year | 8.80 | 9.20 | 8.25 | 7.49 |
| | (29.42) | (29.63) | (28.03) | (27.05) |
| Cumul. days in strong sanctions in the last year - never sanctioned | 0.89 | 0.88 | 0.90 | 0.91 |
| Cumul. days in strong sanctions in the last year - once (3 months) | 0.09 | 0.10 | 0.09 | 0.08 |
| Cumul. days in strong sanctions in the last year - more than once | 0.02 | 0.02 | 0.02 | 0.01 |
| Panel C: Last job characteristics | | | | |
| Type of last job - contributory employment | 0.62 | 0.63 | 0.59 | 0.59 |
| Type of last job - minor employment | 0.32 | 0.31 | 0.37 | 0.36 |
| Type of last job - vocational training | 0.03 | 0.02 | 0.02 | 0.02 |
| Type of last job - no last job | 0.03 | 0.03 | 0.02 | 0.03 |
| Last occupation - agriculture, forestry, gardening | 0.05 | 0.04 | 0.05 | 0.05 |
| Last occupation - manufacturing | 0.13 | 0.13 | 0.10 | 0.12 |
| Last occupation - manufacturing engineering | 0.08 | 0.08 | 0.08 | 0.08 |
| Last occupation - construction | 0.20 | 0.19 | 0.17 | 0.21 |
| Last occupation - grocery, hospitality | 0.08 | 0.08 | 0.09 | 0.08 |
| Last occupation - healthcare | 0.01 | 0.01 | 0.01 | 0.01 |
| Last occupation - humanities, arts | 0.01 | 0.01 | 0.02 | 0.02 |
| Last occupation - trading | 0.06 | 0.07 | 0.08 | 0.07 |
| Last occupation - business management and organization | 0.03 | 0.03 | 0.04 | 0.03 |
| Last occupation - service | 0.03 | 0.03 | 0.04 | 0.04 |
| Last occupation - security, logistic, transport | 0.23 | 0.25 | 0.25 | 0.22 |
| Last occupation - cleaning | 0.03 | 0.04 | 0.04 | 0.04 |
| Last occupation - no last job / missing | 0.05 | 0.05 | 0.04 | 0.06 |
| Last occupational level - assistant | 0.39 | 0.42 | 0.38 | 0.36 |
| Last occupational level - specialist | 0.49 | 0.48 | 0.49 | 0.50 |
| Last occupational level - expert | 0.04 | 0.03 | 0.04 | 0.04 |
| Last occupational level - professional | 0.03 | 0.03 | 0.04 | 0.04 |
| Last occupational level - no last job / missing | 0.05 | 0.05 | 0.04 | 0.06 |
| Last job industry - agriculture, forestry, fishing, mining, manufacturing, energy, water supply | 0.11 | 0.10 | 0.09 | 0.10 |
| Last job industry - construction | 0.11 | 0.10 | 0.11 | 0.12 |
| Last job industry - trade, car sales and maintenance | 0.11 | 0.10 | 0.13 | 0.11 |
| Last job industry - hospitality | 0.08 | 0.08 | 0.11 | 0.09 |
| Last job industry - transport and postal services, telecommunication | 0.08 | 0.08 | 0.09 | 0.08 |
| Last job industry - financial services, real estate, renting out property, services for companies | 0.32 | 0.37 | 0.33 | 0.30 |
| Last job industry - public administration, defense, social security agencies, education, health, and social work | 0.09 | 0.09 | 0.06 | 0.10 |
| Last job industry - other services | 0.06 | 0.06 | 0.06 | 0.07 |
| Last job industry - no last job / missing | 0.03 | 0.03 | 0.02 | 0.04 |
| Last job working time - full-time | 0.61 | 0.61 | 0.58 | 0.57 |
| Last job working time - part-time | 0.36 | 0.36 | 0.40 | 0.40 |
| Last job working time - no last job | 0.03 | 0.03 | 0.02 | 0.04 |
| Last job duration - <1 month | 0.14 | 0.16 | 0.13 | 0.13 |
| Last job duration - 1 - <3 months | 0.21 | 0.21 | 0.21 | 0.21 |
| Last job duration - 3 - <6 months | 0.19 | 0.19 | 0.19 | 0.19 |
| Last job duration - 6 - <12 months | 0.19 | 0.18 | 0.20 | 0.20 |
| Last job duration - 12 - <24 months | 0.12 | 0.12 | 0.12 | 0.12 |
| Last job duration - 24 - <36 months | 0.05 | 0.04 | 0.05 | 0.05 |
| Last job duration - 36 - <60 months | 0.04 | 0.03 | 0.04 | 0.04 |
| Last job duration - 60+ months | 0.04 | 0.04 | 0.05 | 0.04 |
| Last job duration - no last job | 0.03 | 0.03 | 0.02 | 0.03 |
| Last daily real wage (in Euro) | 32.11 | 31.16 | 32.59 | 30.87 |
| | (41.06) | (27.26) | (30.68) | (32.96) |
| Panel D: Labor market status in December 2004 | | | | |
| Dec 2004 - unemployment insurance receipt | 0.09 | 0.10 | 0.11 | 0.09 |
| Dec 2004 - unemployment assistance receipt | 0.35 | 0.35 | 0.32 | 0.37 |
| Dec 2004 - registered unemployment | 0.48 | 0.50 | 0.46 | 0.50 |
| Dec 2004 - registered jobseeker not unemployed | 0.13 | 0.14 | 0.12 | 0.12 |
| Dec 2004 - participation in any ALMP | 0.11 | 0.11 | 0.11 | 0.10 |
| Dec 2004 - contributory employment | 0.24 | 0.23 | 0.26 | 0.21 |
| Dec 2004 - minor employment | 0.08 | 0.08 | 0.09 | 0.08 |
| Panel E: Household level | | | | |
| HH with members aged 18-24 years (1=yes) | 0.08 | 0.08 | 0.08 | 0.08 |
| HH with members aged 25-34 years (1=yes) | 0.44 | 0.47 | 0.42 | 0.38 |
| HH with members aged 35-44 years (1=yes) | 0.37 | 0.37 | 0.40 | 0.38 |
| HH with members aged 45-54 years (1=yes) | 0.30 | 0.27 | 0.32 | 0.36 |
| HH with members aged 55-64 years (1=yes) | 0.01 | 0.01 | 0.00 | 0.01 |
| # of own children aged <3 years | 0.11 | 0.13 | 0.13 | 0.10 |
| # of own children aged 3-5 years | 0.11 | 0.12 | 0.13 | 0.10 |
| # of own children aged 6-9 years | 0.13 | 0.13 | 0.15 | 0.12 |
| # of own children aged 10-12 years | 0.08 | 0.08 | 0.09 | 0.08 |
| # of own children aged 13-14 years | 0.05 | 0.04 | 0.06 | 0.05 |
| # of own children aged 15-17 years | 0.06 | 0.07 | 0.08 | 0.07 |
| HH equiv. UB II income (in prices of 2010) | 677.95 | 697.92 | 719.50 | 717.34 |
| | (266.08) | (240.21) | (253.37) | (226.84) |
| HH receives income from dependent employment | 0.15 | 0.15 | 0.21 | 0.18 |
| HH receives income from self-employment | 0.01 | 0.01 | 0.01 | 0.01 |
| HH receives UB I | 0.05 | 0.06 | 0.06 | 0.04 |
| HH receives income from child support | 0.01 | 0.01 | 0.01 | 0.01 |
| HH receives income from alimony | 0.00 | 0.00 | 0.00 | 0.00 |
| HH receives pension or housing assistance | 0.01 | 0.01 | 0.01 | 0.01 |
| HH receives income from other sources | 0.03 | 0.03 | 0.03 | 0.03 |
| HH - no information UB II income found | 0.05 | 0.03 | 0.02 | 0.01 |
| Panel F: Partner characteristics | | | | |
| Partner's age (in years) | 12.19 | 12.39 | 13.85 | 12.91 |
| | (17.83) | (17.47) | (18.26) | (18.42) |
| Partner's nationality - Germany | 0.23 | 0.24 | 0.21 | 0.24 |
| Partner's nationality - EU (w/o Germany, with former YUG) | 0.02 | 0.02 | 0.03 | 0.02 |
| Partner's nationality - Europe rest (w/o former YUG) | 0.01 | 0.01 | 0.02 | 0.01 |
| Partner's nationality - Turkey | 0.04 | 0.04 | 0.06 | 0.04 |
| Partner's nationality - former Soviet Union (w/o EU members) | 0.01 | 0.02 | 0.02 | 0.01 |
| Partner's nationality - other countries | 0.03 | 0.03 | 0.05 | 0.03 |
| Partner's nationality - missing | 0.66 | 0.65 | 0.62 | 0.65 |
| Partner's continent - Germany | 0.23 | 0.24 | 0.21 | 0.24 |
| Partner's continent - Europe | 0.08 | 0.08 | 0.12 | 0.08 |
| Partner's continent - Africa | 0.01 | 0.01 | 0.02 | 0.01 |
| Partner's continent - America | 0.00 | 0.00 | 0.00 | 0.00 |
| Partner's continent - Asia / Oceania | 0.02 | 0.02 | 0.03 | 0.02 |
| Partner's continent - stateless / unknown | 0.00 | 0.00 | 0.00 | 0.00 |
| Partner's continent - missing | 0.66 | 0.65 | 0.62 | 0.65 |
| Partner's education - no schooling diploma | 0.08 | 0.07 | 0.11 | 0.08 |
| Partner's education - secondary school | 0.13 | 0.14 | 0.14 | 0.13 |
| Partner's education - general certificate of secondary education | 0.08 | 0.09 | 0.07 | 0.09 |
| Partner's education - advanced technical college entrance qualification | 0.01 | 0.01 | 0.01 | 0.01 |
| Partner's education - high school | 0.02 | 0.02 | 0.02 | 0.02 |
| Partner's education - missing | 0.69 | 0.68 | 0.65 | 0.68 |
| Partner's vocational degree - no vocational / academic degree | 0.21 | 0.22 | 0.27 | 0.21 |
| Partner's vocational degree - vocational degree | 0.11 | 0.12 | 0.09 | 0.12 |
| Partner's vocational degree - academic degree | 0.01 | 0.01 | 0.01 | 0.01 |
| Partner's vocational degree - missing | 0.67 | 0.66 | 0.63 | 0.66 |
| Partner's marital status - single | 0.03 | 0.03 | 0.02 | 0.03 |
| Partner's marital status - married | 0.24 | 0.24 | 0.30 | 0.25 |
| Partner's marital status - no longer married | 0.02 | 0.02 | 0.02 | 0.02 |
| Partner's marital status - cohabitation | 0.06 | 0.07 | 0.05 | 0.06 |
| Partner's marital status - missing | 0.66 | 0.64 | 0.61 | 0.65 |
| Partner's disability status (1= yes) | 0.01 | 0.01 | 0.01 | 0.01 |
| Partner's disability status (1= missing) | 0.01 | 0.01 | 0.02 | 0.01 |
| Partner's last job type - contributory employment | 0.13 | 0.14 | 0.14 | 0.14 |
| Partner's last job type - minor employment | 0.12 | 0.12 | 0.14 | 0.12 |
| Partner's last job type - vocational training | 0.01 | 0.01 | 0.01 | 0.01 |
| Partner's last job type - no last job | 0.74 | 0.73 | 0.71 | 0.73 |
| Partner's last occupation - agriculture, forestry, gardening | 0.01 | 0.01 | 0.01 | 0.01 |
| Partner's last occupation - manufacturing | 0.02 | 0.02 | 0.01 | 0.01 |
| Partner's last occupation - manufacturing engineering | 0.01 | 0.01 | 0.01 | 0.01 |
| Partner's last occupation - construction | 0.00 | 0.00 | 0.00 | 0.00 |
| Partner's last occupation - grocery, hospitality | 0.03 | 0.04 | 0.03 | 0.03 |
| Partner's last occupation - healthcare | 0.02 | 0.02 | 0.02 | 0.02 |
| Partner's last occupation - humanities, arts | 0.02 | 0.02 | 0.02 | 0.02 |
| Partner's last occupation - trading | 0.04 | 0.05 | 0.05 | 0.04 |
| Partner's last occupation - business management and organization | 0.02 | 0.02 | 0.02 | 0.02 |
| Partner's last occupation - service | 0.01 | 0.01 | 0.01 | 0.01 |
| Partner's last occupation - security, logistic, transport | 0.02 | 0.02 | 0.02 | 0.02 |
| Partner's last occupation - cleaning | 0.03 | 0.03 | 0.04 | 0.04 |
| Partner's last occupation - no last job / missing | 0.79 | 0.77 | 0.77 | 0.77 |
| Partner's last occupational level - assistant | 0.09 | 0.10 | 0.11 | 0.10 |
| Partner's last occupational level - specialist | 0.11 | 0.12 | 0.11 | 0.11 |
| Partner's last occupational level - expert | 0.01 | 0.01 | 0.01 | 0.01 |
| Partner's last occupational level - professional | 0.01 | 0.01 | 0.01 | 0.01 |
| Partner's last occupational level - no last job / missing | 0.79 | 0.77 | 0.77 | 0.77 |
| Partner's last job industry - agriculture, forestry, fishing, mining, manufacturing, energy, water supply | 0.02 | 0.02 | 0.02 | 0.02 |
| Partner's last job industry - construction | 0.01 | 0.00 | 0.00 | 0.00 |
| Partner's last job industry - trade, car sales and maintenance | 0.04 | 0.05 | 0.05 | 0.05 |
| Partner's last job industry - hospitality | 0.03 | 0.04 | 0.04 | 0.03 |
| Partner's last job industry - transport and postal services, telecommunication | 0.01 | 0.01 | 0.01 | 0.01 |
| Partner's last job industry - financial services, real estate, renting out property, services for companies | 0.08 | 0.08 | 0.10 | 0.08 |
| Partner's last job industry - public administration, defense, social security agencies, education, health, and social work | 0.04 | 0.05 | 0.04 | 0.05 |
| Partner's last job industry - other services | 0.03 | 0.03 | 0.02 | 0.03 |
| Partner's last job industry - no last job / missing | 0.74 | 0.73 | 0.71 | 0.73 |
| Partner's last job working time - full-time | 0.09 | 0.10 | 0.10 | 0.10 |
| Partner's last job working time - part-time | 0.17 | 0.17 | 0.20 | 0.18 |
| Partner's last job working time - no last job | 0.74 | 0.73 | 0.71 | 0.73 |
| Partner's last job duration - <1 month | 0.03 | 0.03 | 0.03 | 0.03 |
| Partner's last job duration - 1-<3 months | 0.04 | 0.05 | 0.05 | 0.05 |
| Partner's last job duration - 3 -<6 months | 0.04 | 0.05 | 0.05 | 0.05 |
| Partner's last job duration - 6 -<12 months | 0.05 | 0.05 | 0.06 | 0.06 |
| Partner's last job duration - 12-<24 months | 0.04 | 0.04 | 0.04 | 0.04 |
| Partner's last job duration - 24-<36 months | 0.02 | 0.02 | 0.02 | 0.02 |
| Partner's last job duration - 36-<60 months | 0.02 | 0.02 | 0.02 | 0.02 |
| Partner's last job duration - 60+ months | 0.02 | 0.02 | 0.02 | 0.02 |
| Partner's last job duration - no last job / missing | 0.74 | 0.73 | 0.71 | 0.73 |
| Partner's last daily wage (in Euro) | 5.55 | 5.82 | 6.43 | 6.05 |
| | (13.59) | (14.52) | (14.72) | (16.04) |
| Partner's cumul. days in unemployment in the last year | 46.21 | 45.29 | 50.62 | 51.47 |
| | (105.83) | (104.25) | (110.41) | (112.58) |
| Partner's cumul. days in unemployment in the last 5 years | 215.30 | 216.78 | 232.63 | 242.83 |
| | (439.36) | (432.42) | (448.30) | (471.53) |
| Partner's cumul. days in job search in the last year | 71.41 | 73.44 | 81.10 | 81.17 |
| | (135.45) | (136.62) | (141.84) | (143.67) |
| Partner's cumul. days in job search in the last 5 years | 316.15 | 324.87 | 346.16 | 357.47 |
| | (569.83) | (566.76) | (579.86) | (606.36) |
| Partner's cumul. days in UB II in the last year | 102.94 | 110.50 | 120.29 | 112.61 |
| | (157.44) | (160.71) | (164.00) | (162.77) |
| Partner's cumul. days in UB II in the last 5 years | 425.53 | 455.29 | 485.31 | 470.71 |
| | (690.33) | (706.39) | (711.56) | (721.03) |
| Partner's days since last regular employment | 206.06 | 229.29 | 217.26 | 225.54 |
| | (647.00) | (676.46) | (655.97) | (678.57) |
| Partner's cumul. days in regular employment in the last year | 14.03 | 16.17 | 18.29 | 16.62 |
| | (64.18) | (68.58) | (72.53) | (70.29) |
| Partner's cumul. days in regular employment in the last 5 years | 81.26 | 90.38 | 92.95 | 87.45 |
| | (296.22) | (311.29) | (308.86) | (310.39) |
| Partner in contributory employment at sampling date | 0.04 | 0.04 | 0.05 | 0.05 |
| Partner in minor employment at sampling date | 0.05 | 0.05 | 0.06 | 0.05 |
| Panel G: District-level information | | | | |
| Unemployment rate (in %) | 10.39 | 10.49 | 10.15 | 10.99 |
| | (3.49) | (3.13) | (3.01) | (3.54) |
| Long-term unemployment rate (in %) | 3.78 | 3.75 | 3.99 | 3.97 |
| | (1.76) | (1.56) | (1.63) | (1.69) |
| Long-term unemployment stock | 8,371.45 | 8,989.26 | 15,501.27 | 14,755.62 |
| | (15,703.18) | (15,233.79) | (18,970.25) | (24,556.80) |
| Unemployment rate of welfare recipients (in %) | 7.32 | 7.42 | 7.46 | 7.90 |
| | (3.13) | (2.88) | (2.89) | (3.26) |
| District vacancy-unemployment ratio | 0.08 | 0.09 | 0.10 | 0.08 |
| | (0.04) | (0.05) | (0.05) | (0.04) |
| Panel H: Information at the job center level | | | | |
| Employees in job center (JC) | 334.49 | 407.95 | 628.75 | 417.02 |
| | (330.57) | (417.74) | (452.81) | (400.36) |
| Share of JC employees in Market and Integration | 0.43 | 0.43 | 0.45 | 0.42 |
| | (0.10) | (0.10) | (0.10) | (0.08) |
| Share of JC employees in Benefits Administration | 0.40 | 0.40 | 0.38 | 0.40 |
| | (0.10) | (0.09) | (0.10) | (0.08) |
| Share of JC employees of female JC employees | 0.68 | 0.68 | 0.66 | 0.68 |
| | (0.10) | (0.10) | (0.07) | (0.09) |
| Share of JC employees being civil servants | 0.17 | 0.17 | 0.21 | 0.17 |
| | (0.08) | (0.08) | (0.08) | (0.08) |
| Share of JC employees on fixed-term contract | 0.23 | 0.23 | 0.24 | 0.22 |
| | (0.07) | (0.06) | (0.05) | (0.07) |
| Share of JC employees on fixed-term contract among employees in Market and Integration | 0.21 | 0.22 | 0.24 | 0.21 |
| | (0.09) | (0.08) | (0.07) | (0.09) |
| Client-staff ratio | 65.77 | 66.73 | 67.80 | 67.06 |
|  | (7.11) | (6.64) | (7.20) | (7.67) |
| Client-staff ratio among employees in Market and Integration | 158.76 (28.54) | 159.58 (28.11) | 155.90 (26.81) | 162.41 (26.19) |
| JC 2009/q4 - people with at least 1 sanction / UB II recipients stock (25-54 years) | 2.62 (0.93) | 2.65 (0.82) | 2.64 (0.79) | 2.60 (0.83) |
| JC 2009/q4 - people with complete sanction (no UB II) / UB II recipients stock (25-54 years) | 0.11 (0.12) | 0.11 (0.11) | 0.09 (0.10) | 0.10 (0.11) |
| JC 2009/q4 - sanction intensity due to failure in reporting (25-54 years) | 0.71 (0.26) | 0.71 (0.27) | 0.71 (0.23) | 0.71 (0.26) |
| JC 2009/q4 - sanction intensity due to violations of duties (25-54 years) | 0.54 (0.28) | 0.54 (0.23) | 0.53 (0.23) | 0.51 (0.24) |
| JC 2009/q4 - inflow into classroom SAI / stock of UB II jobseekers (25-54 years) | 1.69 (1.00) | 1.79 (1.12) | 1.63 (1.06) | 1.30 (0.89) |
| JC 2009/q4 - inflow into in-firm SAI / stock of UB II jobseekers (25-54 years) | 0.41 (0.20) | 0.42 (0.20) | 0.30 (0.23) | 0.38 (0.20) |
| JC 2009/q4 - inflow into further vocational training / stock of UB II recipients (25-54 years) | 0.51 (0.33) | 0.52 (0.31) | 0.35 (0.28) | 0.54 (0.33) |
| JC 2009/q4 - inflow into wage subsidies / stock of UB II jobseekers (25-54 years) | 0.24 (0.10) | 0.25 (0.10) | 0.21 (0.09) | 0.24 (0.10) |
| JC 2009/q4 - inflow into One-Euro-Jobs / stock of UB II jobseekers (25-54 years) | 1.11 (0.56) | 1.10 (0.58) | 0.95 (0.48) | 1.11 (0.61) |
| JC type - Cities west, average labor market situation (LMS), high GDP, high rate of long term unemployed | 0.10 | 0.10 | 0.48 | 0.12 |
| JC type - Cities west, above average LMS, high GDP | 0.06 | 0.05 | 0.06 | 0.04 |
| JC type - Cities west, below average LMS, very high rate of long-term unemployed | 0.17 | 0.15 | 0.14 | 0.22 |
| JC type - Cities, mainly east, bad LMS, very high rate of long-term unemployed | 0.06 | 0.11 | 0.03 | 0.09 |
| JC type - Predominantly urban, west, average LMS, high rate of long-term unemployed | 0.10 | 0.20 | 0.06 | 0.12 |
| JC type - rural, west, average LMS | 0.13 | 0.10 | 0.09 | 0.12 |
| JC type - Predominantly urban, west and east, below average LMS | 0.07 | 0.05 | 0.02 | 0.05 |
| JC type - rural, west, good LMS, high seasonal dynamic | 0.02 | 0.02 | 0.03 | 0.02 |
| JC type - rural, west, very good LMS, seasonal dynamic, very low rate of long-term unemployed | 0.03 | 0.02 | 0.02 | 0.02 |
| JC type - rural, west, very good LMS, low average rate of long-term unemployed | 0.08 | 0.08 | 0.03 | 0.06 |
| JC type - predominantly rural, east, bad LMS, low GDP | 0.10 | 0.09 | 0.03 | 0.10 |
| JC type - predominantly rural, east, very bad LMS, very low GDP, high average rate of long-term unemployed | 0.07 | 0.03 | 0.01 | 0.05 |

Note. − Means of the covariates. Values in parentheses are the standard deviations.

*Table O-D.2: Summary of covariates, women*

| | Women | | | |
|---|---|---|---|---|
| Variable | Job-training | Reducing impediments | Placement services | Non-participants |
| Panel A: Sociodemographic characteristics | | | | |
| Age at sampling date (in years) | 38.90 | 38.53 | 39.62 | 39.86 |
| | (8.48) | (8.27) | (8.30) | (8.41) |
| Region (west=0, east=1) | 0.31 | 0.28 | 0.16 | 0.33 |
| Foreigner | 0.18 | 0.19 | 0.24 | 0.22 |
| Nationality - Germany | 0.82 | 0.81 | 0.76 | 0.78 |
| Nationality - EU (w/o Germany, w/o former YUG) | 0.05 | 0.05 | 0.06 | 0.05 |
| Nationality - Europe rest (w/o former YUG) | 0.02 | 0.02 | 0.03 | 0.02 |
| Nationality - Turkey | 0.05 | 0.05 | 0.07 | 0.07 |
| Nationality - former Soviet Union (without EU members) | 0.03 | 0.03 | 0.03 | 0.03 |
| Nationality - other countries | 0.04 | 0.05 | 0.05 | 0.05 |
| Continent - Germany | 0.82 | 0.81 | 0.76 | 0.78 |
| Continent - Europe | 0.13 | 0.14 | 0.19 | 0.16 |
| Continent - Africa | 0.01 | 0.01 | 0.02 | 0.01 |
| Continent - America | 0.00 | 0.00 | 0.00 | 0.00 |
| Continent - Asia / Oceania | 0.03 | 0.03 | 0.03 | 0.04 |
| Continent - stateless / unknown | 0.00 | 0.00 | 0.00 | 0.00 |
| Education - no schooling diploma | 0.13 | 0.11 | 0.13 | 0.17 |
| Education - secondary school | 0.42 | 0.42 | 0.42 | 0.40 |
| Education - general certificate of secondary education | 0.31 | 0.33 | 0.27 | 0.30 |
| Education - advanced technical college entrance qualification | 0.03 | 0.04 | 0.05 | 0.03 |
| Education - high school | 0.08 | 0.07 | 0.10 | 0.07 |
| Education - missing values | 0.03 | 0.03 | 0.02 | 0.03 |
| No vocational / academic degree | 0.48 | 0.49 | 0.52 | 0.53 |
| Vocational degree | 0.46 | 0.47 | 0.43 | 0.42 |
| Academic degree | 0.04 | 0.04 | 0.05 | 0.04 |
| Vocational degree - missing values | 0.01 | 0.01 | 0.01 | 0.01 |
| Family status - single | 0.31 | 0.31 | 0.30 | 0.28 |
| Family status - married | 0.27 | 0.27 | 0.27 | 0.32 |
| Family status - no longer married | 0.34 | 0.34 | 0.37 | 0.33 |
| Family status - cohabitation | 0.08 | 0.08 | 0.06 | 0.07 |
| Marital status - unmarried | 0.31 | 0.31 | 0.30 | 0.28 |
| Marital status - married | 0.27 | 0.27 | 0.27 | 0.32 |
| Marital status - widowed | 0.02 | 0.01 | 0.01 | 0.02 |
| Marital status - divorced | 0.19 | 0.19 | 0.21 | 0.18 |
| Marital status - separated | 0.14 | 0.14 | 0.14 | 0.13 |
| Marital status - cohabitation | 0.08 | 0.08 | 0.06 | 0.07 |
| BG type - single | 0.30 | 0.28 | 0.32 | 0.27 |
| BG type - single with adult aged <25 years | 0.02 | 0.01 | 0.01 | 0.01 |
| BG type - single with children aged <18 years | 0.28 | 0.31 | 0.29 | 0.27 |
| BG type - single with children aged <18 years and adult aged <25 years | 0.03 | 0.03 | 0.03 | 0.03 |
| BG type - couple w/o children | 0.10 | 0.09 | 0.10 | 0.11 |
| BG type - couple w/o children, but with adult aged <25 years | 0.02 | 0.02 | 0.02 | 0.03 |
| BG type - couple with children aged <18 years | 0.18 | 0.19 | 0.17 | 0.20 |
| BG type - couple with children aged <18 years and adult aged <25 years | 0.03 | 0.03 | 0.03 | 0.04 |
| BG type - other | 0.03 | 0.04 | 0.04 | 0.04 |
| Pregnancy - neither sample member nor partner | 1.00 | 1.00 | 1.00 | 0.98 |
| Pregnancy - sample member | 0.00 | 0.00 | 0.00 | 0.02 |
| Pregnancy - partner | 0.00 | 0.00 | 0.00 | 0.00 |
| Pregnancy - missing value sample member | 0.00 | 0.00 | 0.00 | 0.00 |
| Pregnancy - missing value partner | 0.00 | 0.00 | 0.00 | 0.00 |
| Partner living in HH (1=yes) | 0.32 | 0.32 | 0.29 | 0.36 |
| Child in HH (1=yes) | 0.53 | 0.55 | 0.52 | 0.54 |
| Own child in HH (1=yes) | 0.53 | 0.55 | 0.52 | 0.54 |
| Profile of unemployed - Market | 0.03 | 0.02 | 0.05 | 0.02 |
| Profile of unemployed - Activation | 0.02 | 0.02 | 0.03 | 0.01 |
| Profile of unemployed - Support | 0.20 | 0.22 | 0.27 | 0.18 |
| Profile of unemployed - Development | 0.42 | 0.45 | 0.39 | 0.40 |
| Profile of unemployed - Stabilization | 0.08 | 0.08 | 0.06 | 0.08 |
| Profile of unemployed - Assistance | 0.17 | 0.14 | 0.14 | 0.23 |
| Profile of unemployed - integrated, but dependent on help | 0.02 | 0.01 | 0.02 | 0.02 |
| Profile of unemployed - not specified | 0.02 | 0.02 | 0.02 | 0.02 |
| Profile of unemployed - missing | 0.03 | 0.02 | 0.01 | 0.01 |
| Profile of unemployed - assignment not required | 0.03 | 0.02 | 0.02 | 0.03 |
| Panel B: Labor market history | | | | |
| Cumul. days in unemployment in the last year | 252.02 (119.31) | 259.49 (111.88) | 265.57 (108.79) | 277.59 (106.73) |
| Cumul. days in unemployment in the last year - 1 month | 0.06 | 0.04 | 0.03 | 0.02 |
| Cumul. days in unemployment in the last year - >1-2 months | 0.05 | 0.04 | 0.03 | 0.03 |
| Cumul. days in unemployment in the last year - >2-3 months | 0.05 | 0.05 | 0.04 | 0.04 |
| Cumul. days in unemployment in the last year - >3-6 months | 0.15 | 0.15 | 0.15 | 0.13 |
| Cumul. days in unemployment in the last year - >6-9 months | 0.15 | 0.16 | 0.18 | 0.15 |
| Cumul. days in unemployment in the last year - >9-12 months | 0.54 | 0.57 | 0.58 | 0.63 |
| Cumul. days in unemployment in the last 2 years | 441.19 (236.28) | 450.95 (221.00) | 466.83 (221.70) | 491.66 (219.74) |
| Cumul. days in unemployment in the last 5 years | 944.92 (541.93) | 959.85 (516.84) | 978.67 (520.62) | 1,038.15 (523.84) |
| Cumul. days in unemployment in the last 5 years - 0-6 months | 0.10 | 0.07 | 0.07 | 0.06 |
| Cumul. days in unemployment in the last 5 years - >6-12 months | 0.09 | 0.08 | 0.08 | 0.07 |
| Cumul. days in unemployment in the last 5 years - >12-24 months | 0.20 | 0.23 | 0.21 | 0.19 |
| Cumul. days in unemployment in the last 5 years - >24-36 months | 0.19 | 0.20 | 0.20 | 0.19 |
| Cumul. days in unemployment in the last 5 years - >36-48 months | 0.20 | 0.21 | 0.21 | 0.22 |
| Cumul. days in unemployment in the last 5 years - >48 months | 0.22 | 0.21 | 0.23 | 0.27 |
| Cumul. days in job search in the last year | 303.23 (107.27) | 315.71 (94.87) | 319.77 (88.48) | 323.04 (86.41) |
| Cumul. days in job search in the last 2 years | 554.79 (236.90) | 576.97 (217.52) | 581.59 (211.43) | 597.05 (203.61) |
| Cumul. days in job search in the last 5 years | 1,227.30 (577.52) | 1,260.54 (550.21) | 1,258.27 (546.96) | 1,303.11 (533.31) |
| Cumul. days in UB II in the last year | 310.40 (106.32) | 321.71 (92.54) | 323.03 (89.25) | 332.03 (79.45) |
| Cumul. days in UB II in the last year - 1 month | 0.05 | 0.03 | 0.02 | 0.01 |
| Cumul. days in UB II in the last year - >1-2 months | 0.03 | 0.02 | 0.02 | 0.01 |
| Cumul. days in UB II in the last year - >2-6 months | 0.07 | 0.07 | 0.07 | 0.06 |
| Cumul. days in UB II in the last year - >6-9 months | 0.05 | 0.06 | 0.06 | 0.05 |
| Cumul. days in UB II in the last year - >9-12 months | 0.80 | 0.83 | 0.83 | 0.87 |
| Cumul. days in UB II in the last 2 years | 589.52 (231.79) | 610.14 (209.77) | 607.51 (209.67) | 632.58 (187.83) |
| Cumul. days in UB II in the last 5 years | 1,316.60 (600.14) | 1,352.04 (570.81) | 1,332.04 (574.50) | 1,403.04 (531.32) |
| Cumul. days in UB II in the last 5 years - 0-6 months | 0.09 | 0.06 | 0.06 | 0.04 |
| Cumul. days in UB II in the last 5 years - >6-12 months | 0.04 | 0.05 | 0.05 | 0.04 |
| Cumul. days in UB II in the last 5 years - >12-24 months | 0.08 | 0.08 | 0.09 | 0.08 |
| Cumul. days in UB II in the last 5 years - >24-36 months | 0.09 | 0.09 | 0.10 | 0.09 |
| Cumul. days in UB II in the last 5 years - >36-48 months | 0.12 | 0.12 | 0.12 | 0.13 |
| Cumul. days in UB II in the last 5 years - >48 months | 0.58 | 0.60 | 0.58 | 0.63 |
| Experience in UB I (1=yes) | 0.05 | 0.04 | 0.04 | 0.03 |
| Cumul. days in UB I in the last year | 18.36 (61.63) | 17.97 (60.68) | 19.09 (62.36) | 13.21 (52.26) |
| Cumul. days in UB I in the last 2 years | 31.71 (90.09) | 31.72 (90.43) | 35.12 (94.68) | 25.37 (82.06) |
| Cumul. days in UB I in the last 5 years | 76.65 (150.29) | 79.97 (153.04) | 86.42 (160.45) | 67.77 (143.51) |
| Days since last employment | 1,973.28 (2,116.50) | 1,952.10 (2,118.96) | 1,896.75 (2,164.42) | 2,115.16 (2,287.01) |
| Cumul. days in regular employment in the last year | 17.89 (58.44) | 16.34 (53.71) | 19.71 (56.83) | 12.01 (45.80) |
| Cumul. days in regular employment in the last 5 years | 174.49 (346.47) | 169.05 (335.20) | 196.01 (347.01) | 137.26 (298.46) |
| Cumul. days in contributory employment in the last year | 20.43 (61.74) | 18.90 (57.28) | 21.77 (59.41) | 14.46 (50.17) |
| Cumul. days in contributory employment in the last 2 years | 58.33 (134.15) | 54.95 (127.53) | 62.43 (130.30) | 42.79 (110.70) |
| Cumul. days in contributory employment in the last 5 years | 176.62 (338.42) | 168.63 (325.82) | 196.66 (339.41) | 141.74 (292.74) |
| Cumul. days in vocational training in the last year | 0.82 (13.20) | 0.82 (12.93) | 0.90 (13.31) | 0.57 (10.38) |
| Cumul. days in vocational training in the last 2 years | 2.91 (35.22) | 2.90 (34.63) | 2.39 (32.15) | 2.14 (29.83) |
| Cumul. days in vocational training in the last 5 years | 14.86 (101.50) | 15.83 (108.09) | 12.74 (92.37) | 11.13 (88.90) |
| Cumul. days in minor employment in the last year | 66.65 (124.72) | 62.50 (120.95) | 88.05 (139.04) | 82.89 (136.94) |
| Cumul. days in minor employment in the last 2 years | 129.72 (229.59) | 121.73 (220.20) | 165.15 (251.89) | 153.37 (248.13) |
| Cumul. days in minor employment in the last 5 years | 284.96 (471.93) | 270.29 (454.65) | 348.13 (504.82) | 318.90 (500.52) |
| Cumul. days in bridging allowance in the last 5 years | 0.56 (9.83) | 0.53 (9.51) | 0.75 (11.55) | 0.58 (9.86) |
| Cumul. days in One-Euro-Jobs in the last year | 13.38 (47.84) | 15.91 (52.15) | 12.65 (46.68) | 12.51 (45.51) |
| Cumul. days in One-Euro-Jobs in the last 2 years | 30.79 (84.23) | 37.15 (91.88) | 28.26 (79.15) | 29.99 (83.50) |
| Cumul. days in One-Euro-Jobs in the last 5 years | 72.51 (152.58) | 81.30 (159.03) | 60.13 (134.12) | 69.38 (151.64) |
| | 2.05 | 2.25 | 1.64 | 2.13 |
| Cumul. days in subsidized public employment in the last year | (19.17) | (20.71) | (17.48) | (19.99) |
| Cumul. days in subsidized public employment in the last 2 years | 5.31 (35.00) | 4.98 (34.81) | 3.54 (28.93) | 5.17 (35.61) |
| Cumul. days in subsidized public employment in the last 5 years | 11.72 (55.93) | 10.18 (51.81) | 8.07 (45.31) | 11.06 (55.50) |
| Cumul. days in subsidized employment in the last year | 1.27 (14.50) | 1.04 (11.43) | 1.28 (13.00) | 0.82 (10.36) |
| Cumul. days in subsidized employment in the last 5 years | 5.49 (33.59) | 5.52 (32.30) | 5.69 (34.38) | 4.81 (31.30) |
| Experience in further vocational training (1=yes) | 0.26 | 0.27 | 0.27 | 0.24 |
| Cumul. days in further vocational training in the last year | 6.69 (32.25) | 6.06 (29.12) | 8.61 (36.10) | 5.85 (29.62) |
| Cumul. days in further vocational training in the last 5 years | 23.77 (88.84) | 24.80 (85.95) | 26.51 (89.35) | 21.03 (82.79) |
| Experience in in-firm SAI (1=yes) | 0.02 | 0.03 | 0.03 | 0.02 |
| Cumul. days in in-firm SAI in the last year | 0.24 (2.06) | 0.30 (2.35) | 0.32 (2.60) | 0.24 (2.21) |
| Cumul. days in in-firm SAI in the last 2 years | 0.24 (2.06) | 0.30 (2.35) | 0.32 (2.60) | 0.24 (2.21) |
| Experience in classroom SAI (1=yes) | 0.11 | 0.14 | 0.09 | 0.06 |
| Cumul. days in classroom SAI in the last year | 4.46 (21.89) | 6.15 (22.96) | 5.75 (25.35) | 2.84 (17.28) |
| Cumul. days in classroom SAI in the last 2 years | 4.46 (21.89) | 6.15 (22.96) | 5.75 (25.35) | 2.84 (17.28) |
| Experience in *job-training* (*JT*) (1=yes) | 0.07 | 0.05 | 0.03 | 0.02 |
| Experience in *reducing impediments* (*RIM*) (1=yes) | 0.02 | 0.07 | 0.02 | 0.02 |
| Cumul. days in SAI in the last year | 4.65 (21.77) | 6.40 (22.84) | 6.06 (25.45) | 3.07 (17.38) |
| Cumul. days in SAI in the last 2 years | 4.65 (21.77) | 6.40 (22.84) | 6.06 (25.45) | 3.07 (17.38) |
| Experience in in-firm training (1=yes) | 0.13 | 0.14 | 0.11 | 0.11 |
| Cumul. days in in-firm training in the last year | 0.26 (2.23) | 0.35 (2.83) | 0.33 (2.70) | 0.26 (2.37) |
| Cumul. days in in-firm training in the last 5 years | 1.96 (9.19) | 2.04 (9.01) | 1.73 (8.82) | 1.51 (7.78) |
| Experience in classroom training (1=yes) | 0.50 | 0.56 | 0.47 | 0.40 |
| Cumul. days in classroom training in the last year | 6.75 (24.67) | 9.55 (26.53) | 9.02 (28.64) | 4.53 (19.94) |
| Cumul. days in classroom training in the last 5 years | 17.69 (37.71) | 22.62 (40.93) | 17.63 (37.65) | 12.11 (30.55) |
| Ever received mild sanction (1=yes) | 0.09 | 0.09 | 0.09 | 0.10 |
| Cumul. days in mild sanctions in the last year | 4.06 (21.58) | 3.81 (20.53) | 4.19 (21.13) | 4.38 (22.42) |
| Cumul. days in mild sanctions in the last year - never sanctioned | 0.96 | 0.96 | 0.95 | 0.95 |
| Cumul. days in mild sanctions in the last year - once (3 months) | 0.04 | 0.03 | 0.04 | 0.04 |
| Cumul. days in mild sanctions in the last year - more than once | 0.01 | 0.01 | 0.01 | 0.01 |
| Ever received strong sanction (1=yes) | 0.10 | 0.10 | 0.09 | 0.09 |
| Cumul. days in strong sanctions in the last year | 3.66 (18.87) | 3.84 (18.79) | 3.91 (19.14) | 2.97 (16.84) |
| Cumul. days in strong sanctions in the last year - never sanctioned | 0.95 | 0.95 | 0.95 | 0.96 |
| Cumul. days in strong sanctions in the last year - once (3 months) | 0.04 | 0.05 | 0.04 | 0.03 |
| Cumul. days in strong sanctions in the last year - more than once | 0.01 | 0.01 | 0.01 | 0.00 |

Panel C: Last job characteristics

| | | | | |
|---|---|---|---|---|
| Type of last job - contributory employment | 0.41 | 0.42 | 0.40 | 0.36 |
| Type of last job - minor employment | 0.47 | 0.47 | 0.52 | 0.51 |
| Type of last job - vocational training | 0.03 | 0.03 | 0.02 | 0.02 |
| Type of last job - no last job | 0.09 | 0.09 | 0.07 | 0.11 |
| Last occupation - agriculture, forestry, gardening | 0.04 | 0.03 | 0.04 | 0.04 |
| Last occupation - manufacturing | 0.05 | 0.04 | 0.04 | 0.05 |
| Last occupation - manufacturing engineering | 0.03 | 0.03 | 0.02 | 0.02 |
| Last occupation - construction | 0.01 | 0.01 | 0.01 | 0.01 |
| Last occupation - grocery, hospitality | 0.14 | 0.13 | 0.13 | 0.12 |
| Last occupation - healthcare | 0.06 | 0.06 | 0.06 | 0.06 |
| Last occupation - humanities, arts | 0.05 | 0.05 | 0.05 | 0.05 |
| Last occupation - trading | 0.15 | 0.16 | 0.18 | 0.14 |
| Last occupation - business management and organization | 0.08 | 0.10 | 0.11 | 0.08 |
| Last occupation - service | 0.03 | 0.04 | 0.05 | 0.04 |
| Last occupation - security, logistic, transport | 0.09 | 0.08 | 0.07 | 0.07 |
| Last occupation - cleaning | 0.11 | 0.09 | 0.10 | 0.10 |
| Last occupation - no last job / missing | 0.17 | 0.17 | 0.15 | 0.22 |
| Last occupational level - assistant | 0.37 | 0.34 | 0.34 | 0.33 |
| Last occupational level - specialist | 0.41 | 0.43 | 0.43 | 0.40 |
| Last occupational level - expert | 0.03 | 0.03 | 0.04 | 0.03 |
| Last occupational level - professional | 0.03 | 0.02 | 0.04 | 0.03 |
| Last occupational level - no last job / missing | 0.17 | 0.17 | 0.15 | 0.22 |
| Last job industry - agriculture, forestry, fishing, mining, manufacturing, energy, water supply | 0.08 | 0.07 | 0.07 | 0.07 |
| Last job industry - construction | 0.01 | 0.01 | 0.01 | 0.01 |
| Last job industry - trade, car sales and maintenance | 0.15 | 0.17 | 0.18 | 0.15 |
| Last job industry - hospitality | 0.13 | 0.12 | 0.13 | 0.13 |
| Last job industry - transport and postal services, telecommunication | 0.04 | 0.04 | 0.04 | 0.03 |
| Last job industry - financial services, real estate, renting out property, services for companies | 0.26 | 0.27 | 0.28 | 0.25 |
| Last job industry - public administration, defense, social security agencies, education, health, and social work | 0.15 | 0.13 | 0.12 | 0.14 |
| Last job industry - other services | 0.10 | 0.10 | 0.11 | 0.11 |
| Last job industry - no last job / missing | 0.17 | 0.17 | 0.15 | 0.22 |
| Last job working time - full-time | 0.33 | 0.33 | 0.33 | 0.30 |
| Last job working time - part-time | 0.58 | 0.58 | 0.60 | 0.60 |
| Last job working time - no last job | 0.09 | 0.09 | 0.07 | 0.11 |
| Last job duration - <1 month | 0.10 | 0.10 | 0.09 | 0.09 |
| Last job duration - 1 - <3 months | 0.16 | 0.18 | 0.17 | 0.16 |
| Last job duration - 3 - <6 months | 0.16 | 0.16 | 0.16 | 0.16 |
| Last job duration - 6 - <12 months | 0.20 | 0.20 | 0.19 | 0.19 |
| Last job duration - 12 - <24 months | 0.13 | 0.13 | 0.14 | 0.13 |
| Last job duration - 24 - <36 months | 0.06 | 0.06 | 0.07 | 0.06 |
| Last job duration - 36 - <60 months | 0.05 | 0.04 | 0.06 | 0.05 |
| Last job duration - 60+ months | 0.05 | 0.05 | 0.06 | 0.05 |
| Last job duration - no last job | 0.09 | 0.09 | 0.07 | 0.11 |
| Last daily real wage (in Euro) | 18.93 | 19.33 | 20.13 | 17.60 |
| | (20.64) | (31.30) | (22.78) | (21.44) |
| Panel D: Labor market status in December 2004 | | | | |
| Dec 2004 - unemployment insurance receipt | 0.06 | 0.07 | 0.07 | 0.06 |
| Dec 2004 - unemployment assistance receipt | 0.23 | 0.23 | 0.21 | 0.23 |
| Dec 2004 - registered unemployment | 0.36 | 0.36 | 0.35 | 0.36 |
| Dec 2004 - registered jobseeker not unemployed | 0.11 | 0.10 | 0.10 | 0.10 |
| Dec 2004 - participation in any ALMP | 0.07 | 0.07 | 0.08 | 0.06 |
| Dec 2004 - contributory employment | 0.16 | 0.16 | 0.18 | 0.14 |
| Dec 2004 - minor employment | 0.14 | 0.14 | 0.16 | 0.14 |
| Panel E: Household level | | | | |
| HH with members aged 18-24 years (1=yes) | 0.11 | 0.11 | 0.10 | 0.13 |
| HH with members aged 25-34 years (1=yes) | 0.36 | 0.37 | 0.33 | 0.32 |
| HH with members aged 35-44 years (1=yes) | 0.39 | 0.41 | 0.40 | 0.40 |
| HH with members aged 45-54 years (1=yes) | 0.35 | 0.33 | 0.38 | 0.39 |
| HH with members aged 55-64 years (1=yes) | 0.02 | 0.02 | 0.02 | 0.03 |
| # of own children aged <3 years | 0.02 | 0.02 | 0.01 | 0.02 |
| # of own children aged 3-5 years | 0.20 | 0.19 | 0.16 | 0.18 |
| # of own children aged 6-9 years | 0.25 | 0.27 | 0.24 | 0.27 |
| # of own children aged 10-12 years | 0.17 | 0.20 | 0.20 | 0.19 |
| # of own children aged 13-14 years | 0.10 | 0.11 | 0.10 | 0.12 |
| # of own children aged 15-17 years | 0.14 | 0.14 | 0.14 | 0.16 |
| HH equiv. UB II income (in prices of 2010) | 649.11 | 664.22 | 695.43 | 668.09 |
| | (255.10) | (240.89) | (238.07) | (232.37) |
| HH receives income from dependent employment | 0.21 | 0.20 | 0.30 | 0.27 |
| HH receives income from self-employment | 0.01 | 0.01 | 0.01 | 0.01 |
| HH receives UB I | 0.04 | 0.04 | 0.04 | 0.03 |
| HH receives income from child support | 0.06 | 0.06 | 0.04 | 0.05 |
| HH receives income from alimony | 0.01 | 0.02 | 0.02 | 0.01 |
| HH receives pension or housing assistance | 0.02 | 0.02 | 0.01 | 0.02 |
| HH receives income from other sources | 0.04 | 0.04 | 0.04 | 0.04 |
| HH - no information UB II income found | 0.03 | 0.02 | 0.01 | 0.01 |
| Panel F: Partner characteristics | | | | |
| Partner's age (in years) | 13.65 | 13.67 | 12.74 | 15.71 |
| | (20.52) | (20.55) | (20.29) | (21.62) |
| Partner's nationality - Germany | 0.23 | 0.23 | 0.19 | 0.25 |
| Partner's nationality - EU (w/o Germany, with former YUG) | 0.01 | 0.01 | 0.02 | 0.02 |
| Partner's nationality - Europe rest (w/o former YUG) | 0.01 | 0.01 | 0.01 | 0.01 |
| Partner's nationality - Turkey | 0.03 | 0.03 | 0.04 | 0.05 |
| Partner's nationality - former Soviet Union (w/o EU members) | 0.01 | 0.02 | 0.01 | 0.01 |
| Partner's nationality - other countries | 0.02 | 0.02 | 0.02 | 0.02 |
| Partner's nationality - missing | 0.68 | 0.68 | 0.71 | 0.64 |
| Partner's continent - Germany | 0.24 | 0.23 | 0.19 | 0.25 |
| Partner's continent - Europe | 0.06 | 0.06 | 0.08 | 0.08 |
| Partner's continent - Africa | 0.01 | 0.01 | 0.01 | 0.01 |
| Partner's continent - America | 0.00 | 0.00 | 0.00 | 0.00 |
| Partner's continent - Asia / Oceania | 0.02 | 0.02 | 0.02 | 0.02 |
| Partner's continent - stateless / unknown | 0.00 | 0.00 | 0.00 | 0.00 |
| Partner's continent - missing | 0.68 | 0.68 | 0.71 | 0.64 |
| Partner's education - no schooling diploma | 0.05 | 0.05 | 0.05 | 0.07 |
| Partner's education - secondary school | 0.15 | 0.15 | 0.14 | 0.17 |
| Partner's education - general certificate of secondary education | 0.08 | 0.08 | 0.07 | 0.08 |
| Partner's education - advanced technical college entrance qualification | 0.01 | 0.01 | 0.01 | 0.01 |
| Partner's education - high school | 0.02 | 0.02 | 0.03 | 0.02 |
| Partner's education - missing | 0.69 | 0.69 | 0.71 | 0.65 |
| Partner's vocational degree - no vocational / academic degree | 0.17 | 0.17 | 0.17 | 0.20 |
| Partner's vocational degree - vocational degree | 0.14 | 0.14 | 0.11 | 0.14 |
| Partner's vocational degree - academic degree | 0.01 | 0.01 | 0.01 | 0.01 |
| Partner's vocational degree - missing | 0.68 | 0.68 | 0.71 | 0.64 |
| Partner's marital status - single | 0.02 | 0.02 | 0.01 | 0.02 |
| Partner's marital status - married | 0.23 | 0.23 | 0.22 | 0.27 |
| Partner's marital status - no longer married | 0.01 | 0.01 | 0.01 | 0.01 |
| Partner's marital status - cohabitation | 0.06 | 0.06 | 0.05 | 0.06 |
| Partner's marital status - missing | 0.68 | 0.68 | 0.71 | 0.64 |
| Partner's disability status (1= yes) | 0.01 | 0.01 | 0.01 | 0.02 |
| Partner's disability status (1= missing) | 0.00 | 0.00 | 0.00 | 0.00 |
| Partner's last job type - contributory employment | 0.21 | 0.21 | 0.19 | 0.23 |
| Partner's last job type - minor employment | 0.09 | 0.09 | 0.09 | 0.11 |
| Partner's last job type - vocational training | 0.00 | 0.00 | 0.00 | 0.00 |
| Partner's last job type - no last job | 0.69 | 0.69 | 0.72 | 0.66 |
| Partner's last occupation - agriculture, forestry, gardening | 0.02 | 0.02 | 0.01 | 0.02 |
| Partner's last occupation - manufacturing | 0.04 | 0.04 | 0.03 | 0.04 |
| Partner's last occupation - manufacturing engineering | 0.03 | 0.02 | 0.02 | 0.03 |
| Partner's last occupation - construction | 0.07 | 0.07 | 0.05 | 0.07 |
| Partner's last occupation - grocery, hospitality | 0.03 | 0.03 | 0.03 | 0.03 |
| Partner's last occupation - healthcare | 0.00 | 0.00 | 0.00 | 0.00 |
| Partner's last occupation - humanities, arts | 0.00 | 0.00 | 0.00 | 0.00 |
| Partner's last occupation - trading | 0.02 | 0.02 | 0.02 | 0.02 |
| Partner's last occupation - business management and organization | 0.01 | 0.01 | 0.01 | 0.01 |
| Partner's last occupation - service | 0.01 | 0.01 | 0.01 | 0.01 |
| Partner's last occupation - security, logistic, transport | 0.07 | 0.08 | 0.08 | 0.08 |
| Partner's last occupation - cleaning | 0.01 | 0.01 | 0.01 | 0.01 |
| Partner's last occupation - no last job / missing | 0.70 | 0.70 | 0.72 | 0.66 |
| Partner's last occupational level - assistant | 0.12 | 0.11 | 0.11 | 0.13 |
| Partner's last occupational level - specialist | 0.16 | 0.16 | 0.14 | 0.18 |
| Partner's last occupational level - expert | 0.01 | 0.01 | 0.01 | 0.01 |
| Partner's last occupational level - professional | 0.01 | 0.01 | 0.01 | 0.01 |
| Partner's last occupational level - no last job / missing | 0.70 | 0.70 | 0.72 | 0.66 |
| Partner's last job industry - agriculture, forestry, fishing, mining, manufacturing, energy, water supply | 0.04 | 0.03 | 0.03 | 0.04 |
| Partner's last job industry - construction | 0.04 | 0.04 | 0.03 | 0.04 |
| Partner's last job industry - trade, car sales and maintenance | 0.04 | 0.04 | 0.04 | 0.04 |
| Partner's last job industry - hospitality | 0.03 | 0.03 | 0.03 | 0.03 |
| Partner's last job industry - transport and postal services, telecommunication | 0.03 | 0.04 | 0.04 | 0.04 |
| Partner's last job industry - financial services, real estate, renting out property, services for companies | 0.09 | 0.09 | 0.08 | 0.10 |
| Partner's last job industry - public administration, defense, social security agencies, education, health, and social work | 0.03 | 0.03 | 0.02 | 0.03 |
| Partner's last job industry - other services | 0.02 | 0.02 | 0.02 | 0.02 |
| Partner's last job industry - no last job / missing | 0.69 | 0.69 | 0.72 | 0.66 |
| Partner's last job working time - full-time | 0.20 | 0.19 | 0.17 | 0.21 |
| Partner's last job working time - part-time | 0.11 | 0.11 | 0.11 | 0.13 |
| Partner's last job working time - no last job | 0.69 | 0.69 | 0.72 | 0.66 |
| Partner's last job duration - <1 month | 0.03 | 0.03 | 0.03 | 0.03 |
| Partner's last job duration - 1-<3 months | 0.05 | 0.06 | 0.05 | 0.06 |
| Partner's last job duration - 3 -<6 months | 0.05 | 0.06 | 0.05 | 0.06 |
| Partner's last job duration - 6 -<12 months | 0.06 | 0.06 | 0.05 | 0.07 |
| Partner's last job duration - 12-<24 months | 0.04 | 0.04 | 0.04 | 0.05 |
| Partner's last job duration - 24-<36 months | 0.02 | 0.02 | 0.02 | 0.02 |
| Partner's last job duration - 36-<60 months | 0.02 | 0.02 | 0.02 | 0.02 |
| Partner's last job duration - 60+ months | 0.02 | 0.03 | 0.02 | 0.03 |
| Partner's last job duration - no last job / missing | 0.69 | 0.69 | 0.72 | 0.66 |
| Partner's last daily wage (in Euro) | 10.65 | 10.42 | 9.61 | 11.72 |
| | (22.08) | (21.63) | (22.87) | (26.03) |
| Partner's cumul. days in unemployment in the last year | 58.90 | 57.57 | 55.44 | 66.62 |
| | (117.56) | (116.80) | (115.77) | (124.52) |
| Partner's cumul. days in unemployment in the last 5 years | 279.47 | 274.63 | 265.78 | 325.35 |
| | (508.51) | (504.21) | (508.23) | (544.66) |
| Partner's cumul. days in job search in the last year | 95.45 | 95.81 | 89.26 | 108.90 |
| | (153.97) | (154.59) | (150.82) | (160.63) |
| Partner's cumul. days in job search in the last 5 years | 415.16 | 417.81 | 381.86 | 479.31 |
| | (676.82) | (678.85) | (657.37) | (710.20) |
| Partner's cumul. days in UB II in the last year | 98.35 | 100.20 | 95.01 | 115.97 |
| | (155.99) | (157.22) | (154.92) | (164.28) |
| Partner's cumul. days in UB II in the last 5 years | 423.58 | 429.07 | 392.51 | 494.45 |
| | (701.36) | (702.79) | (681.94) | (735.45) |
| Partner's days since last regular employment | 269.94 | 265.87 | 247.78 | 320.91 |
| | (717.15) | (709.22) | (674.94) | (778.54) |
| Partner's cumul. days in regular employment in the last year | 21.59 | 22.54 | 18.64 | 22.74 |
| | (76.89) | (78.79) | (71.44) | (79.12) |
| Partner's cumul. days in regular employment in the last 5 years | 130.03 | 134.56 | 115.92 | 134.47 |
| | (359.61) | (371.04) | (342.83) | (363.88) |
| Partner in contributory employment at sampling date | 0.06 | 0.06 | 0.05 | 0.06 |
| Partner in minor employment at sampling date | 0.04 | 0.05 | 0.04 | 0.05 |
| Panel G: District-level information | | | | |
| Unemployment rate (in %) | 10.35 | 10.44 | 10.19 | 10.79 |
| | (3.50) | (3.26) | (3.07) | (3.56) |
| Long-term unemployment rate (in %) | 3.73 | 3.73 | 4.00 | 3.89 |
| | (1.75) | (1.61) | (1.65) | (1.72) |
| Long-term unemployment stock | 7,476.41 | 8,421.65 | 15,364.40 | 13,517.08 |
| | (14,024.04) | (14,693.48) | (18,602.57) | (23,215.80) |
| Unemployment rate of welfare recipients (in %) | 7.26 | 7.37 | 7.50 | 7.71 |
| | (3.11) | (2.97) | (2.92) | (3.28) |
| District vacancy-unemployment ratio | 0.08 | 0.09 | 0.10 | 0.08 |
| | (0.05) | (0.05) | (0.05) | (0.04) |
| Panel H: Information at the job center level | | | | |
| Employees in job center (JC) | 321.19 | 379.58 | 626.56 | 400.14 |
| | (322.15) | (379.49) | (462.92) | (389.11) |
| Share of JC employees in Market and Integration | 0.42 | 0.43 | 0.44 | 0.42 |
| | (0.10) | (0.11) | (0.09) | (0.08) |
| Share of JC employees in Benefits Administration | 0.40 | 0.40 | 0.39 | 0.41 |
| | (0.09) | (0.10) | (0.09) | (0.08) |
| Share of JC employees of female JC employees | 0.69 | 0.68 | 0.67 | 0.68 |
| | (0.10) | (0.10) | (0.07) | (0.09) |
| Share of JC employees being civil servants | 0.16 | 0.17 | 0.20 | 0.17 |
| | (0.08) | (0.08) | (0.08) | (0.08) |
| Share of JC employees on fixed-term contract | 0.23 | 0.23 | 0.24 | 0.22 |
| | (0.07) | (0.06) | (0.06) | (0.07) |
| Share of JC employees on fixed-term contract among employees in Market and Integration | 0.21 | 0.22 | 0.25 | 0.21 |
| | (0.09) | (0.08) | (0.08) | (0.09) |
| Client-staff ratio | 65.51 | (66.80) | (67.77) | 66.86 |
| | (7.26) | (6.91) | (7.14) | (7.64) |
| | 159.60 | 159.82 | 157.69 | 162.13 |
| Client-staff ratio among employees in Market and Integration | (27.80) | (29.39) | (26.29) | (26.69) |
| JC 2009/q4 - people with at least 1 sanction / UB II recipients stock (25-54 years) | 2.63 (0.95) | 2.65 (0.82) | 2.65 (0.83) | 2.62 (0.85) |
| JC 2009/q4 - people with complete sanction (no UB II) / UB II recipients stock (25-54 years) | 0.12 (0.12) | 0.11 (0.11) | 0.09 (0.11) | 0.11 (0.12) |
| JC 2009/q4 - sanction intensity due to failure in reporting (25-54 years) | 0.71 (0.27) | 0.72 (0.27) | 0.72 (0.24) | 0.71 (0.26) |
| JC 2009/q4 - sanction intensity due to violations of duties (25-54 years) | 0.55 (0.28) | 0.55 (0.23) | 0.53 (0.23) | 0.52 (0.25) |
| JC 2009/q4 - inflow into classroom SAI / stock of UB II jobseekers (25-54 years) | 1.70 (0.98) | 1.80 (1.14) | 1.61 (1.05) | 1.32 (0.90) |
| JC 2009/q4 - inflow into in-firm SAI / stock of UB II jobseekers (25-54 years) | 0.42 (0.19) | 0.42 (0.20) | 0.32 (0.25) | 0.38 (0.20) |
| JC 2009/q4 - inflow into further vocational training / stock of UB II recipients (25-54 years) | 0.51 (0.33) | 0.51 (0.30) | 0.34 (0.30) | 0.53 (0.33) |
| JC 2009/q4 - inflow into wage subsidies / stock of UB II jobseekers (25-54 years) | 0.25 (0.11) | 0.25 (0.09) | 0.21 (0.10) | 0.24 (0.10) |
| JC 2009/q4 - inflow into One-Euro-Jobs / stock of UB II jobseekers (25-54 years) | 1.11 (0.57) | 1.06 (0.58) | 0.92 (0.47) | 1.11 (0.61) |
| JC type - Cities west, average labor market situation (LMS), high GDP, high rate of long term unemployed | 0.09 | 0.11 | 0.46 | 0.11 |
| JC type - Cities west, above average LMS, high GDP | 0.06 | 0.05 | 0.05 | 0.04 |
| JC type - Cities west, below average LMS, very high rate of long-term unemployed | 0.15 | 0.14 | 0.12 | 0.21 |
| JC type - Cities, mainly east, bad LMS, very high rate of long-term unemployed | 0.07 | 0.11 | 0.04 | 0.08 |
| JC type - Predominantly urban, west, average LMS, high rate of long-term unemployed | 0.11 | 0.18 | 0.06 | 0.12 |
| JC type - rural, west, average LMS | 0.13 | 0.08 | 0.10 | 0.13 |
| JC type - Predominantly urban, west and east, below average LMS | 0.07 | 0.05 | 0.02 | 0.05 |
| JC type - rural, west, good LMS, high seasonal dynamic | 0.03 | 0.02 | 0.05 | 0.02 |
| JC type - rural, west, very good LMS, seasonal dynamic, very low rate of long-term unemployed | 0.03 | 0.02 | 0.02 | 0.03 |
| JC type - rural, west, very good LMS, low average rate of long-term unemployed | 0.08 | 0.11 | 0.03 | 0.07 |
| JC type - predominantly rural, east, bad LMS, low GDP | 0.11 | 0.10 | 0.05 | 0.10 |
| JC type - predominantly rural, east, very bad LMS, very low GDP, high average rate of long-term unemployed | 0.08 | 0.04 | 0.02 | 0.05 |

Note. – Means of the covariates. Values in parentheses are the standard deviations.

# Online Appendix O-E: Full descriptive statistics of clusters

*Table O-E.1: Descriptive statistics of clusters based on k-means clustering, men, full table*

| Cluster | Least beneficial | 2 | 3 | 4 | Most beneficial |
|---|---|---|---|---|---|
| Share of observations (in %) | 13 | 30 | 23 | 22 | 12 |
| JT vs. NP | 27 | 29 | 31 | 40 | 68 |
| RIM vs. NP | 14 | 23 | 41 | 53 | 39 |
| PS vs. NP | 7 | 53 | 35 | 61 | 55 |
| Region (west=0, east=1) | 0.26 | 0.44 | 0.36 | 0.26 | 0.26 |
| Foreigner | 0.26 | 0.11 | 0.22 | 0.26 | 0.22 |
| Days in regular employment (last 5 years) | 704 | 156 | 203 | 135 | 201 |
| Days since last employment | 453 | 2413 | 1663 | 1775 | 1758 |
| Client-staff ratio in job centers | 161 | 159 | 164 | 163 | 163 |
| Sanction intensity in job centers due to violations of duties (in percent) | 0.58 | 0.50 | 0.51 | 0.50 | 0.53 |
| Sanction intensity in job centers due to failure in reporting (in percent) | 0.75 | 0.71 | 0.72 | 0.70 | 0.72 |
| District unemployment rate | 9.76 | 11.36 | 11.05 | 10.87 | 10.52 |
| District unemployment rate of welfare recipients | 6.77 | 8.14 | 7.95 | 7.93 | 7.59 |
| No vocational / academic degree | 0.47 | 0.41 | 0.56 | 0.60 | 0.47 |
| Vocational degree | 0.49 | 0.56 | 0.41 | 0.37 | 0.42 |
| Academic degree | 0.03 | 0.03 | 0.02 | 0.03 | 0.09 |
| Education - No schooling diploma | 0.12 | 0.11 | 0.17 | 0.16 | 0.11 |
| Education - Secondary school | 0.50 | 0.47 | 0.51 | 0.50 | 0.39 |
| Education - General certificate of secondary education | 0.27 | 0.31 | 0.24 | 0.21 | 0.23 |
| Education - Advanced technical college entrance qualification | 0.04 | 0.03 | 0.03 | 0.04 | 0.08 |
| Education - High school | 0.07 | 0.07 | 0.06 | 0.07 | 0.15 |
| Nationality - Germany | 0.74 | 0.89 | 0.78 | 0.74 | 0.78 |
| Nationality - European Union | 0.05 | 0.02 | 0.03 | 0.05 | 0.04 |
| Nationality - Rest of Europe | 0.05 | 0.01 | 0.02 | 0.03 | 0.03 |
| Nationality - Turkey | 0.10 | 0.04 | 0.09 | 0.09 | 0.08 |
| Nationality - Former Soviet Union | 0.02 | 0.01 | 0.02 | 0.02 | 0.02 |
| Nationality - Rest of the world | 0.06 | 0.03 | 0.05 | 0.07 | 0.06 |
| Marital status - unmarried | 0.39 | 0.53 | 0.45 | 0.51 | 0.50 |
| Marital status – married | 0.37 | 0.19 | 0.30 | 0.26 | 0.28 |
| Marital status – widowed | 0.00 | 0.00 | 0.00 | 0.00 | 0.00 |
| Marital status – divorced | 0.09 | 0.15 | 0.08 | 0.09 | 0.08 |
| Marital status – separated | 0.05 | 0.06 | 0.05 | 0.06 | 0.05 |
| Marital status – cohabitation | 0.10 | 0.06 | 0.11 | 0.07 | 0.08 |
| Household – single; no children | 0.48 | 0.72 | 0.54 | 0.62 | 0.59 |
| Household – single; adult children | 0.00 | 0.00 | 0.00 | 0.00 | 0.00 |
| Household – single; child. aged below 18 | 0.01 | 0.02 | 0.02 | 0.02 | 0.01 |
| Household – single; adult and child. <18 | 0.00 | 0.00 | 0.00 | 0.00 | 0.00 |
| Household – couple; no children | 0.10 | 0.08 | 0.11 | 0.09 | 0.10 |
| Household – couple; adult children | 0.02 | 0.02 | 0.02 | 0.02 | 0.01 |
| Household – couple; child. aged below 18 | 0.35 | 0.13 | 0.26 | 0.21 | 0.26 |
| Household – couple; adult and child. < 18 | 0.04 | 0.03 | 0.04 | 0.03 | 0.02 |
| Household – other type | 0.01 | 0.01 | 0.01 | 0.01 | 0.01 |

Note. – NP: non-participation, JT: job-training, RIM: reducing impediments, PS: placement services.

*Table O-E.2: Descriptive statistics of clusters based on k-means clustering, women, full table*

| Cluster | Least beneficial | 2 | 3 | 4 | Most beneficial |
|---|---|---|---|---|---|
| Share of observations (in %) | 13 | 25 | 18 | 17 | 27 |
| JT vs. NP | 3 | 18 | 29 | 33 | 37 |
| RIM vs. NP | 33 | 37 | 14 | 47 | 46 |
| PS vs. NP | 8 | 28 | 52 | 91 | 49 |
| Region (west=0, east=1) | 0.36 | 0.70 | 0.26 | 0.07 | 0.16 |
| Foreigner | 0.20 | 0.12 | 0.22 | 0.29 | 0.25 |
| Days in regular employment (last 5 years) | 265 | 68 | 174 | 229 | 84 |
| Days since last employment | 1798 | 2610 | 1655 | 1794 | 2244 |
| Client-staff ratio in job centers | 164 | 165 | 162 | 159 | 159 |
| Sanction intensity in job centers due to violations of duties (in percent) | 0.45 | 0.42 | 0.48 | 0.62 | 0.62 |
| Sanction intensity in job centers due to failure in reporting (in percent) | 0.63 | 0.66 | 0.69 | 0.76 | 0.79 |
| District unemployment rate | 11.33 | 13.09 | 10.88 | 8.58 | 9.57 |
| District unemployment rate of welfare recipients | 8.21 | 9.58 | 7.89 | 5.78 | 6.69 |
| No vocational / academic degree | 0.49 | 0.38 | 0.47 | 0.61 | 0.65 |
| Vocational degree | 0.47 | 0.57 | 0.45 | 0.34 | 0.31 |
| Academic degree | 0.03 | 0.04 | 0.07 | 0.03 | 0.02 |
| Education - No schooling diploma | 0.15 | 0.12 | 0.11 | 0.18 | 0.23 |
| Education - Secondary school | 0.40 | 0.36 | 0.32 | 0.47 | 0.45 |
| Education - General certificate of secondary education | 0.33 | 0.42 | 0.35 | 0.21 | 0.21 |
| Education - Advanced technical college entrance qualification | 0.03 | 0.03 | 0.06 | 0.03 | 0.03 |
| Education - High school | 0.07 | 0.07 | 0.12 | 0.06 | 0.05 |
| Nationality - Germany | 0.80 | 0.88 | 0.78 | 0.71 | 0.75 |
| Nationality - European Union | 0.05 | 0.03 | 0.05 | 0.07 | 0.05 |
| Nationality - Rest of Europe | 0.02 | 0.01 | 0.02 | 0.04 | 0.02 |
| Nationality - Turkey | 0.06 | 0.03 | 0.06 | 0.10 | 0.09 |
| Nationality - Former Soviet Union | 0.02 | 0.02 | 0.04 | 0.03 | 0.03 |
| Nationality - Rest of the world | 0.05 | 0.03 | 0.06 | 0.06 | 0.06 |
| Marital status - unmarried | 0.27 | 0.37 | 0.38 | 0.17 | 0.22 |
| Marital status – married | 0.31 | 0.17 | 0.21 | 0.50 | 0.39 |
| Marital status – widowed | 0.01 | 0.02 | 0.01 | 0.01 | 0.02 |
| Marital status – divorced | 0.20 | 0.23 | 0.17 | 0.15 | 0.17 |
| Marital status – separated | 0.12 | 0.13 | 0.17 | 0.11 | 0.13 |
| Marital status – cohabitation | 0.09 | 0.08 | 0.07 | 0.06 | 0.07 |
| Household – single; no children | 0.26 | 0.38 | 0.34 | 0.18 | 0.22 |
| Household – single; adult children | 0.01 | 0.02 | 0.01 | 0.01 | 0.01 |
| Household – single; child. aged below 18 | 0.27 | 0.31 | 0.34 | 0.18 | 0.26 |
| Household – single; adult and child. <18 | 0.03 | 0.04 | 0.03 | 0.02 | 0.03 |
| Household – couple; no children | 0.14 | 0.08 | 0.06 | 0.17 | 0.11 |
| Household – couple; adult children | 0.04 | 0.02 | 0.01 | 0.04 | 0.03 |
| Household – couple; child. aged below 18 | 0.18 | 0.09 | 0.16 | 0.30 | 0.25 |
| Household – couple; adult and child. <18 | 0.04 | 0.02 | 0.02 | 0.07 | 0.06 |
| Household – other type | 0.03 | 0.05 | 0.03 | 0.03 | 0.03 |

Note. – NP: non-participation, JT: job-training, RIM: reducing impediments, PS: placement services.